\documentclass[10pt,3p]{elsarticle}

\usepackage{amsmath,amsfonts,amssymb}
\usepackage{amsthm}
\usepackage{bm}
\usepackage{graphicx,epstopdf}
\usepackage[position=top,labelfont=normalfont,textfont=normalfont,singlelinecheck=off,justification=raggedright]{subfig}
\usepackage{floatrow}
\usepackage[dvipsnames]{xcolor}
\usepackage{setspace}
\usepackage{enumerate,etaremune}
\usepackage{booktabs}
\usepackage{tabularx,multirow}
\usepackage{framed}
\usepackage{hhline}
\usepackage{url}
\usepackage{import}
\usepackage[unicode,bookmarks=false]{hyperref}
\hypersetup{
  colorlinks,
  citecolor=Blue,linkcolor=Blue,urlcolor=Blue
}

\usepackage{lineno}

\usepackage[textsize=footnotesize,linecolor=red,backgroundcolor=red!25,bordercolor=red]
  {todonotes}

\usepackage{tikz}
\usetikzlibrary{shapes.geometric}
\usetikzlibrary{shapes,arrows}
\usetikzlibrary{plotmarks}

\usetikzlibrary{calc}

\usepackage{physics}

\usepackage{algorithm}
\usepackage{algorithmicx,algpseudocode}
\usepackage{cases}
\newcommand{\eg}{{\it e.g.,}}
\newcommand{\ie}{{\it i.e.,}}

\newcommand{\etal}{{\it et al.}}
\newcommand{\tensor}[1]{\bm{#1}}
\newcommand{\stress}{\sigma}
\newcommand{\strain}{\varepsilon}
\newcommand{\tstress}{\tensor{\stress}}
\newcommand{\tstrain}{\tensor{\strain}}

\newcommand{\pd}{\partial}

\newcommand{\rn}[1]{\uppercase\expandafter{\romannumeral #1\relax}}

\DeclareMathOperator{\symgrad}{\nabla^{s}}

\DeclareMathOperator{\dyadic}{\otimes}
\newcolumntype{L}[1]{>{\raggedright\let\newline\\arraybackslash\hspace{0pt}}m{#1}}
\newcolumntype{C}[1]{>{\centering\let\newline\\arraybackslash\hspace{0pt}}m{#1}}
\newcolumntype{R}[1]{>{\raggedleft\let\newline\\arraybackslash\hspace{0pt}}m{#1}}

\allowdisplaybreaks

\biboptions{sort&compress,square,comma,numbers}
\AtBeginDocument{\hypersetup{citecolor=MidnightBlue,linkcolor=MidnightBlue,urlcolor=MidnightBlue}}

\usepackage{etoolbox}
\makeatletter
\patchcmd{\ps@pprintTitle}% <cmd>
{Preprint submitted to}% <search>
{~}% <replace>
{}{}% <succes><failure>
\makeatother
\begin{document}

\begin{frontmatter}

\title{A phase-field model for quasi-dynamic rupture nucleation and propagation of in-plane faults}

\author[LLNL]{Fan Fei}
\author[UIUC_ME,UIUC_CEE]{Md Shumon Mia}
\author[UIUC_CEE,UIUC_BECKMAN]{Ahmed E. Elbanna\corref{corr}}
\ead{elbanna2@illinois.edu}
\author[KAIST]{Jinhyun Choo\corref{corr}}
\ead{jinhyun.choo@kaist.ac.kr}

\cortext[corr]{Corresponding Authors}
\address[LLNL]{Atmospheric, Earth, and Energy Division, Lawrence Livermore National Laboratory, Livermore, CA, USA}
\address[UIUC_ME]{Department of Mechanical Science and Engineering, University of Illinois at Urbana-Champaign, Urbana, IL, USA}
\address[UIUC_CEE]{Department of Civil and Environmental Engineering, University of Illinois at Urbana-Champaign, Urbana, IL, USA}
\address[UIUC_BECKMAN]{Beckman Institute of Advanced Science and Technology, University of Illinois at Urbana-Champaign, Urbana, IL, USA}
\address[KAIST]{Department of Civil and Environmental Engineering, KAIST, Daejeon, South Korea}

\journal{~}

\begin{abstract}
Computational modeling of faulting processes is an essential tool for understanding earthquake mechanics but remains challenging due to the structural and material complexities of fault zones.
The phase-field method has recently enabled unified modeling of fault propagation and off-fault damage; however, its capability has been restricted to simplified anti-plane settings. 
In this study, we extend the phase-field method to in-plane faulting by introducing two key advancements: (i) the incorporation of enhanced fault kinematics and pressure-dependent shear strength for a more accurate representation of fault behavior, and (ii) a revised fault propagation criterion that explicitly accounts for the coupling between shear strength and normal stress.
The proposed formulation is verified against standard discontinuous approaches to quasi-dynamic fault rupture under in-plane conditions and validated using experimental observations and numerical data on fault nucleation and propagation. 
Simulations incorporating structural complexities and material heterogeneities demonstrate the robustness and versatility of the phase-field model, establishing it as a powerful tool for investigating the interactions between fault zone properties and earthquake processes.
\end{abstract}


\end{frontmatter}

% \linenumbers

% SECTION 1
% ------------------------------------------------------------------------------
\section{Introduction}
\label{sec:intro}
Earthquakes rank among the most extreme natural events, triggered by the sudden release of energy during rapid sliding and growth of frictional discontinuities—commonly referred to as geologic faults in Earth’s crust. 
Despite decades of progress in earthquake simulation, it remains a significant challenge to accurately predict fault movement and propagation, largely due to the complex structural and material properties of fault zones. 
These complexities exert dominant control over fault dynamics, including rupture characteristics such as speed and size~\cite{harris1997effects,dunham2003supershear,poliakov2002dynamic,bhat2004dynamic,ma2015effect,ma2019dynamic}, as well as energy partitioning during seismic events~\cite{sibson1977fault,andrews2005rupture,okubo2019dynamics}. 
Moreover, dynamic ruptures can further alter fault zone properties through processes of damage and healing. This intricate interaction between rupture dynamics and fault zone evolution is central to understanding and modeling earthquake source physics.

Structural and material complexities within fault zones have been extensively documented and analyzed. Immature fault systems typically consist of distributed networks of fractures characterized by irregular structural features such as bends~\cite{mueller1997geomorphic,sathiakumar2021stop}, branches~\cite{kim2004fault,fliss2005fault}, and stepovers~\cite{wesnousky2006predicting,biasi2016steps}. 
These geometric irregularities critically affect rupture dynamics, including the potential for rupture arrest or continuation across varying fault geometries. 
Moreover, they play a key role in fault zone evolution, driving processes such as the nucleation of new faults, the propagation of pre-existing faults, and the coalescence of smaller fault segments.
As fault slip progresses, fault zones transition toward mature systems, forming a localized and simplified fault core that accommodates the majority of shear deformation. However, even in mature fault systems, structural and material heterogeneities persist. 
Fault cores often exhibit non-planar geometries and are surrounded by damage zones with elastic properties distinct from the host rock. 
These damage zones are characterized by secondary faults~\cite{rowe2018geometric}, microcracks~\cite{ben2003characterization,mitchell2009nature,lewis2010diversity}, and altered elastic properties~\cite{ben2003characterization,peng2006temporal,allam2014seismic,zigone2015seismic}, which collectively influence seismic wave behavior through processes such as reflection, dispersion, and diffraction~\cite{lewis2010diversity,yang2015recent,ben2002dynamic,huang2011pulse,zhao2024dynamic}.
Effectively modeling fault slip and growth in such complex systems requires a comprehensive framework capable of capturing the intricate interactions between frictional slip, fault nucleation, fault propagation, and the evolving bulk properties of fault zones.

Fault rupture dynamics have traditionally been modeled using discontinuous approaches, which conceptualize the fault core as a zero-thickness interface characterized by a sharp displacement discontinuity. This sharp-interface representation is rooted in fracture mechanics and supported by field evidence, which shows that fault cores under significant shear typically have negligible thickness---a few centimeters or less~\cite{ben2003characterization}---relative to the fault's overall length and the dominant wavelengths of seismic waves propagating in the far-field.
Among discontinuous approaches, the traction-at-slip-node technique has been widely implemented in finite difference and finite element methods~\cite{andrews2005rupture,templeton2008off,dalguer2007staggered,duan2008inelastic}. This technique explicitly captures frictional slip by separating nodes along the fault interface. 
However, it requires the fault interface to align with element boundaries, limiting its applicability for modeling fault propagation with unknown geometries unless coupled with sophisticated remeshing algorithms.
Advanced alternatives, including the assumed enhanced strain (AES) method~\cite{borja2007continuum,foster2007embedded} and the extended/generalized finite element method (XFEM/GFEM)~\cite{liu2009extended,coon2011nitsche,liu2013extended}, address this limitation by allowing discontinuities to exist within the interior of discretized elements. 
These methods achieve this by enriching the basis functions in the standard Galerkin formulation, enabling the simulation of complex fault geometries including branching and intersections. 
However, their implementation remains computationally and labor-intensive, particularly for cases involving intricate geometric configurations.

Meanwhile, significant efforts have been made to incorporate off-fault damage zones into discontinuous approaches to account for the effects of structural and material complexities. 
A common strategy involves representing the near-fault region as an idealized material inclusion with reduced elastic moduli, termed a low-velocity fault zone (LVFZ)~\cite{ben2002dynamic,huang2014earthquake,ma2015effect}, a hallmark feature of off-fault damage. 
Studies have shown that the predefined width of the LVFZ and the material contrast between the LVFZ and the surrounding homogeneous medium significantly influence seismic characteristics, such as peak slip rates, rupture propagation~\cite{ben2002dynamic}, and the complexity of seismic cycles~\cite{abdelmeguid2019novel,thakur2020effects,nie2022rupture}. 
Consequently, accurate characterization of the geometry and material properties of the LVFZ is essential for reliable predictions of fault source behavior.
Another approach models damage zones through off-fault plasticity~\cite{templeton2008off,dunham2011earthquakea,dunham2011earthquakeb,mia2022spatio,mia2023spectrum}, wherein the accumulation of plastic deformation represents the evolution of off-fault damage and faults.
However, these models are significantly limited by their inability to account for the reduction of elastic moduli in damaged materials. 
This omission prevents them from representing the stiffness reduction during rapid rupture and the subsequent inter-seismic healing processes~\cite{peng2006temporal,ben2019representation}. To address this limitation, continuum damage--breakage models have been proposed~\cite{lyakhovsky2014continuum,lyakhovsky2016dynamic,zhao2024dynamic}, which explicitly incorporate material degradation and healing within earthquake simulations.
Critically, all these approaches treat on-fault frictional response and off-fault damage evolution as decoupled mechanisms~\cite{gabriel2021unified}, rather than addressing them within a unified framework. 
This separation remains a fundamental limitation in capturing the coupled dynamics of faulting processes.

Over the past decade, the phase-field method has emerged as a powerful tool for modeling geologic fractures.
By representing fractures as a diffuse damage field, this approach captures fracture propagation through the evolution of a continuous field governed by principles of fracture mechanics---the same theoretical foundation underpinning discontinuous methods.  
The phase-field method enables the modeling of arbitrary fracture geometries without requiring complex algorithms or enrichment functions. 
Furthermore, like continuum damage models, it inherently incorporates material stiffness degradation as damage progresses. 
This feature makes it particularly well-suited for the unified modeling of frictional slip, fault nucleation, fault propagation, and the evolution of bulk material properties.

The phase-field method has been extended to model frictional geologic discontinuities such as faults. 
Fei and Choo~\cite{fei2020phasea} introduced a phase-field formulation for modeling frictional contact along discontinuities. 
Building on this framework, the same authors developed a phase-field model for frictional shear fractures in quasi-brittle geologic materials~\cite{fei2020phaseb}, deriving governing equations consistent with the fracture mechanics theory proposed by Palmer and Rice~\cite{palmer1973growth}.  
Bryant and Sun~\cite{bryant2021phase} subsequently incorporated rate- and state-dependent friction into the phase-field framework, leveraging concepts established by Fei and Choo~\cite{fei2020phasea}.
Despite these advancements, existing models remain limited in addressing fault rupture processes during earthquakes, as they are constrained to quasi-static conditions and neglect dynamic or quasi-dynamic effects. 
A notable exception is the recent work by Hayek \etal~\cite{hayek2023diffuse}, which introduced a fully dynamic phase-field model for earthquake rupture by combining the diffuse phase-field representation with the classical stress-glut method~\cite{andrews1999test,dalguer2006comparison}. 
However, this model focuses exclusively on planar, non-propagating faults and does not consider fault propagation or off-fault damage evolution. 
Similarly, Gabriel \etal~\cite{gabriel2021unified} proposed a dynamic fault rupture model based on a regularized damage formulation akin to the phase-field approach. 
While this model can capture off-fault shear crack generation, it is derived from damage rheology and cannot fully recover rate- and state-dependent friction, as achieved by classical discontinuous methods.

Recently, Fei \etal~\cite{fei2023phase} introduced the first phase-field formulation for modeling quasi-dynamic rupture and propagation of rate- and state-dependent frictional faults. 
This formulation builds on a phase-field framework for geologic discontinuities~\cite{fei2022phase}, where displacement-jump-based fault kinematics are transformed into a continuous, strain-based representation compatible with phase-field methods for frictional discontinuities. 
A key strength of this approach is its ability to simultaneously model the growth of localized fault slip surfaces and distributed off-fault damage within a unified framework, eliminating the need for remeshing algorithms or separate treatments of damage evolution required in discontinuous approaches.
However, this initial work is restricted to a two-dimensional (2D) anti-plane formulation.
Anti-plane models, commonly used in the study of strike-slip faults~\cite{rice1993spatio,hajarolasvadi2017new,abdelmeguid2019novel}, offer significant numerical efficiency and analytical simplicity by focusing on Mode III ruptures.
Nevertheless, in-plane or three-dimensional (3D) formulations are more suitable for capturing the inherent complexities of natural fault zones, including variable orientations of fault segments~\cite{preuss2019seismic} and diverse en echelon slip surface patterns~\cite{harris1999dynamic,mia2024rupture}.
Unlike anti-plane models, in-plane and 3D formulations allow for variations in mean stress, which create asymmetries in the bulk response to fault slip. 
This distinction is particularly important, as materials in the dilatational quadrants of faults are more susceptible to damage due to these stress variations. 
In contrast, anti-plane models assume constant mean stress, limiting their ability to capture such phenomena.
Developing an in-plane formulation that incorporates the feedback between evolving mean stress and frictional crack growth is therefore essential. 
Such an advancement would enable the simulation of comprehensive fault zone architectures, enhancing the realism and predictive capability of fault zone modeling.

In this work, we propose an advanced in-plane phase-field model that extends the previous anti-plane formulation by incorporating additional kinematic considerations and coupling normal and shear stress components through updated frictional strength and fracture energy.
To reduce the computational demands of fully dynamic simulations, we adopt a quasi-dynamic scheme that approximates inertial effects using a radiation damping term~\cite{rice1993spatio}.
The modeling framework for fault nucleation and propagation under in-plane conditions builds on the anti-plane formulation by partitioning the total potential energy density into strain energy, frictional dissipation, and a viscous dissipation term to approximate radiated energy. 
This approach ensures that the resultant phase-field evolution equation is intrinsically consistent with the earthquake energy budget proposed by Kanamori and Heaton~\cite{kanamori2000microscopic}.
Furthermore, we refine the fault propagation direction criterion by incorporating the coupling between shear strength and normal stress. This enhancement allows the model to capture variations in fault orientations under different magnitudes of confining pressure. Numerical examples under in-plane loading conditions demonstrate the model’s ability to simulate fault nucleation and propagation across diverse structural configurations.

The structure of the paper is as follows. 
Section~\ref{sec:approximation} introduces the in-plane formulation for phase-field modeling of quasi-dynamic, rate- and state-dependent frictional fault slip. 
Section~\ref{sec:formulation} presents the complete phase-field evolution equation for modeling quasi-dynamic fault growth under in-plane conditions. 
To predict the direction of new faults under in-plane loading, the criterion from the phase-field model for quasi-static frictional shear fractures~\cite{fei2020phaseb} is extended to incorporate the effects of normal stress on fault propagation direction.
Section~\ref{sec:verification} verifies the performance of the proposed formulation by modeling in-plane fault rupture, emphasizing the reproduction of the quasi-dynamic rate- and state-dependent frictional responses observed in conventional discontinuous approaches. 
Section~\ref{sec:simulation} further explores the potential of the phase-field method through simulations addressing fault nucleation and propagation under varying structural and material complexities. 
Finally, Section~\ref{sec:closure} concludes the study and outlines directions for future research.

% SECTION 2
% ------------------------------------------------------------------------------
\section{Phase-field approximation of quasi-dynamic fault slip}
\label{sec:approximation}

This section presents the formulation for quasi-dynamic phase-field modeling of rate- and state-dependent frictional fault slip.
To enable in-plane modeling, the formulation incorporates additional fault kinematics and the coupling between normal and shear stresses, building on the original anti-plane version~\cite{fei2023phase}. 

\subsection{Problem statement and phase-field approximation}
Consider a solid domain $\Omega$ with an exterior boundary denoted by $\pd \Omega$.
The exterior boundary is decomposed into a Dirichlet part $\pd_{u} \Omega$, and a Neumann part $\pd_{t} \Omega$, satisfying $\pd_{u} \Omega \cup \pd_{t} \Omega = \pd \Omega$ and $\pd_{u} \Omega \cap \pd_{t} \Omega = \emptyset$. 
The domain also contains a set of sharp interfaces, $\Gamma$, representing fault surfaces. 
The governing equations of momentum balance for the displacement field, $\tensor{u}$, are expressed as (neglecting body forces) 
\begin{align}
    \div \tstress(\tensor{u}) = \rho\ddot{\tensor{u}} \quad & \text{in} \:\: \Omega , \label{eq:momentum-balance} \\ 
	\tstress \cdot \tensor{v} = \hat{\tensor{t}} \quad &\text{on} \:\: \pd_{t} \Omega, \\ 
	\tensor{u} = \hat{\tensor{u}} \quad & \text{on} \:\: \pd_{u} \Omega, \\ 
    \tstress_{\Gamma +} \cdot \tensor{n} = - \tstress_{\Gamma -} \cdot \tensor{n} = \tensor{t}_{\Gamma} \quad &\text{on} \:\: \Gamma ,  \label{eq:stress-continuity}
\end{align}
where $\tstress$ is the stress tensor, $\rho$ is the density, and $\ddot{\tensor{u}}$ is the acceleration. The terms $\hat{\tensor{t}}$ and $\hat{\tensor{u}}$ denote the prescribed traction and displacement on their exterior boundaries, respectively, and $\tensor{v}$ and $\tensor{n}$ denote the unit normal vectors to $\pd_{t} \Omega$ and $\Gamma$, respectively. 
Here, Eq.~\eqref{eq:stress-continuity} enforces the stress continuity constraint across the fault surface, where $\tensor{t}_{\Gamma}$ denotes the traction acting on $\Gamma$.

\begin{figure}
    \centering
    \includegraphics[width=\textwidth]{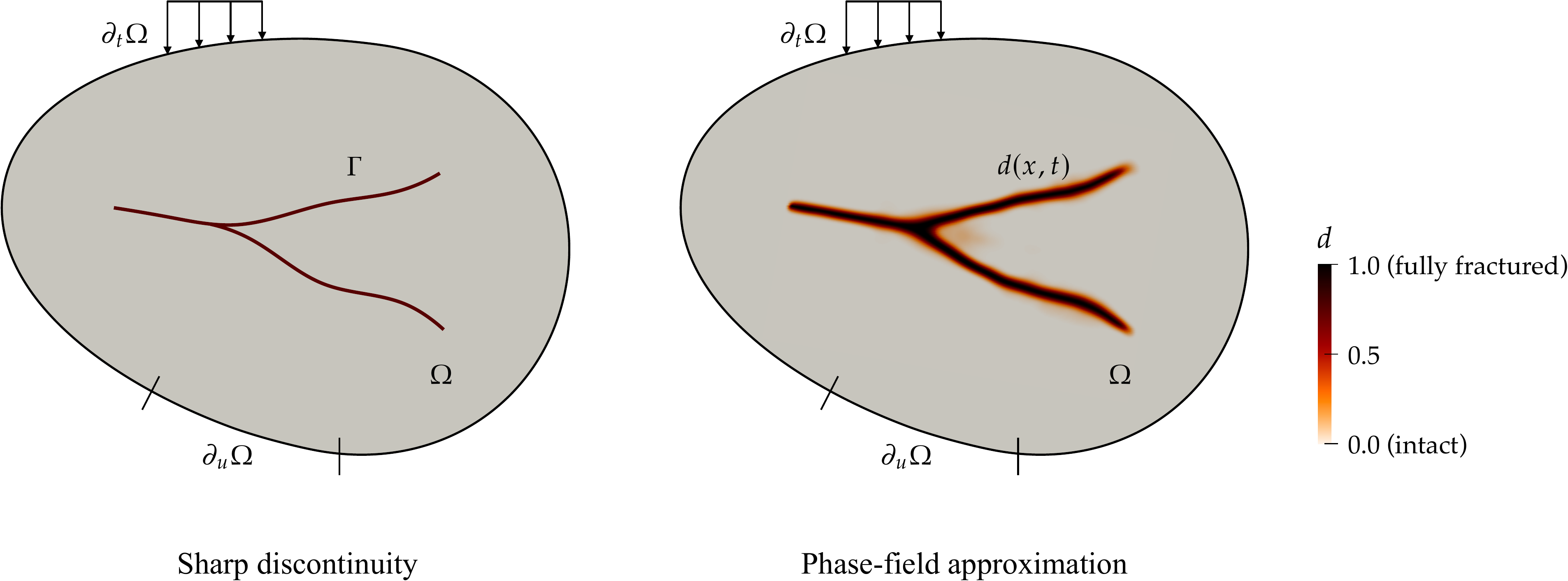}
    \caption{Phase-field approximation of sharp discontinuity. The sharp discontinuity $\Gamma$ on the left is approximated by the diffuse phase-field variable $d$ on the right.}
    \label{fig:pf-approx}
\end{figure}

Figure~\ref{fig:pf-approx} illustrates the phase-field approximation of sharp fault surfaces, $\Gamma$, using a diffusely distributed damage variable, $d \in [0,1]$, where $d = 0$ represents an intact region and $d=1$ represents a fully damaged region.  
In this diffuse representation of $\Gamma$, a crack density function $\Gamma_{d}(d, \grad d)$ is introduced such that
\begin{align}
    \int_\Gamma \dd A = \int_{\Omega} \delta_{\Gamma} (\tensor{x}) \: \dd V \approx \int_\Omega \Gamma_{d} (d, \grad d) \: \dd V, \label{eq:delta-approximation}
\end{align}
where $\delta_{\Gamma}(\tensor{x})$ denotes the Dirac delta function. 
Among various choices for $\Gamma_{d}(d, \grad d)$, we adopt the form proposed by Pham \etal~\cite{pham2011gradient}, given by 
\begin{align}
    \Gamma_{d}(d, \grad d) = \dfrac{3}{8} \left[\dfrac{d}{L} + L \grad d \cdot \grad d \right],  \label{eq:crack-density-function}
\end{align}
where $L$ is the phase-field regularization length that defines the width of the diffuse region. 
Compared to the quadratic form commonly used in standard phase-field models~\cite{miehe2010phase,borden2012phase}, Eq.~\eqref{eq:crack-density-function} enables one to capture bounded peak stress during fracture initiation~\cite{geelen2019phase,fei2020phaseb,fei2023phase}.
This feature makes it particularly well-suited for modeling rocks and rock-like materials, whose fracturing behavior is characterized by strength-dominated failure and post-peak softening. 

After approximating the fault slip surface using the phase-field method, the momentum balance equation~\eqref{eq:momentum-balance} is reformulated as a function of both $\tensor{u}$ and $d$, as
\begin{align}
    \div \tstress(\tensor{u}, d) = \rho\ddot{\tensor{u}} \quad & \text{in} \:\: \Omega . 
    \label{eq:momentum-balance-pf}
\end{align}
The boundary conditions are
\begin{align}
    \tstress \cdot \tensor{v} = \hat{\tensor{t}} \quad &\text{on} \:\: \pd_{t} \Omega, \\ 
    \tensor{u} = \hat{\tensor{u}} \quad & \text{on} \:\: \pd_{u} \Omega. 
\end{align}
Unlike sharp-interface models, the stress continuity constraint becomes unnecessary within the diffuse phase-field representation, as the material properties and stress remain continuous across the diffuse interface. However, this continuity makes it challenging to explicitly represent the surface traction $\tensor{t}_\Gamma$, complicating the incorporation of constitutive responses typical of fractures, such as rate- and state-dependent friction.
To address this issue, we adopt a phase-field modeling framework for frictional contact~\cite{fei2020phasea, fei2023phase}. 
Within this framework, the stress tensor is decomposed into contributions from the bulk material,
$\tstress_{m}$, and the fault interface, $\tstress_{f}$, as
\begin{align}
    \tstress = g(d) \tstress_{m} + [1 - g(d)] \tstress_{f}.
    \label{eq:stress-decomposition}
\end{align}
Here, $g(d)$ is a degradation function that captures the reduction of material stiffness due to damage or fracturing. The degradation function should satisfy
\begin{align}
	g(0) = 1, \:\: g(1) = 0 \:\: \text{and} \:\: g'(d) < 0 \:\: \text{when} \:\: 0<d<1. 
\end{align}
The specific form of $g(d)$ is provided in Section~\ref{sec:formulation}, where the phase-field equation for fault growth modeling is detailed. 
Assuming an isotropic, linear elastic bulk material, the stress in the bulk, $\tstress_{m}$, is calculated as 
\begin{align}
    \tstress_{m} = \mathbb{C}^\mathrm{e} : \tstrain, \:\: \text{with} \:\: \mathbb{C}^\mathrm{e} := K\tensor{1} \dyadic \tensor{1} + 2G \left(\mathbb{I} - \dfrac{1}{3}\tensor{1} \dyadic \tensor{1} \right), 
\end{align}
where $\tstrain := \grad^\text{s} \tensor{u} $ is the strain tensor, $K$ and $G$ are the bulk and shear moduli, respectively, and $\tensor{1}$ and $\mathbb{I}$ are the second and fourth order identity tensors, respectively. 
For the fault interface stress, $\tstress_{f}$, it is formulated to accommodate specific constitutive relations governing $\tensor{t}_\Gamma$. 
The details of these relations are provided in the following.

\subsection{Interface stress formulation through kinematic transformation}
A key challenge in formulating the interface stress lies in the fact that most existing constitutive laws for fault slip (\eg~rate- and state-dependent friction) are based on displacement or velocity jumps, which are not explicitly represented in the continuous phase-field interface.
To address this challenge, we adopt a kinematic transformation strategy introduced by Fei \etal~\cite{fei2022phase}, which recasts the discontinuous, displacement-jump-based constitutive relations into a continuous, strain-based form compatible with the phase-field framework. 
Unlike its application in the anti-plane model, where the transformation is limited to a single displacement scalar, here we apply this approach to the full displacement tensor to accommodate the complexities of fault kinematics in in-plane conditions.

We begin by presenting the original displacement-jump-based kinematics of the fault, decomposing the total displacement into a continuous deformation component, $\bar{\tensor{u}}$, and a displacement jump component, $[\![\tensor{u} ]\!]$, as
\begin{align}
	\tensor{u} = \bar{\tensor{u}} + H_{\Gamma}(\tensor{x})[\![ \tensor{u}]\!],  \label{eq:disp-decompose}
\end{align}
where $H_{\Gamma}(\tensor{x})$ denotes a Heaviside function, defined as
\begin{align}
	H_{\Gamma}(\tensor{x}) = 
	\left \{
	\begin{array}{ll}
		1 & \text{if} \:\: \tensor{x} \in \Omega_{+},   \vspace{0.5em} \\
		0 & \text{if} \:\: \tensor{x} \in \Omega_{-}.
	\end{array}
	\right . 
\end{align}
Given our focus on fault slip, we consider only the tangential component of the displacement jump, neglecting the normal component (i.e., opening or closure) for simplicity.  
Consequently, $[\![ \tensor{u}]\!] $ can be expressed as
\begin{align}
	[\![ \tensor{u} ]\!] = \tensor{m} \zeta_{f} \alpha  , \label{eq:disp-jump-slip}
\end{align}
where $\tensor{m}$ is the unit vector tangential to the fault plane, $\zeta_{f}$ represents the magnitude of slip, and $\alpha = \pm 1$ indicates the slip direction. 
Substituting Eq.~\eqref{eq:disp-jump-slip} into Eq.~\eqref{eq:disp-decompose} and applying the symmetric gradient to both sides yields the strain tensor as 
\begin{align}
	\tstrain := \grad^\mathrm{s} \tensor{u} = \bar{\tstrain} + \dfrac{1}{2} H_{\Gamma}(\tensor{x}) (\tensor{m} \dyadic \grad \zeta_{f} + \grad \zeta_{f} \dyadic \tensor{m}) \alpha  + \dfrac{1}{2} (\tensor{n} \dyadic \tensor{m} + \tensor{m} \dyadic \tensor{n}) \zeta_{f} \delta_{\Gamma} (\tensor{x}) \alpha .  \label{eq:strain-decompose}
\end{align}
In the equation above, $\bar{\tstrain} := \symgrad \bar{\tensor{u}}$ represents the strain associated with continuous deformation, and the Dirac-delta function $\delta_{\Gamma}(\tensor{x})$ arises from the gradient of the Heaviside function, \ie~$\delta_{\Gamma} (\tensor{x}) := \grad H_{\Gamma}(\tensor{x})$. 
Assuming no variation in slip along the fault interface~\cite{regueiro2001plane}, \ie~$\grad \zeta_{f} = \tensor{0}$, and defining the slip direction tensor as  $\tensor{\mathcal{S}} := (\tensor{n} \dyadic \tensor{m} + \tensor{m} \dyadic \tensor{n})/2$, Eq.~\eqref{eq:strain-decompose} simplifies to 
\begin{align}
	\tstrain  = \bar{\tstrain} + \tensor{\mathcal{S}} \zeta_{f} \delta_{\Gamma} (\tensor{x}) \alpha .  \label{eq:strain-decompose-reduced}
\end{align}
To adapt Eq.~\eqref{eq:strain-decompose-reduced} to the phase-field model, we apply the approximation from Eq.~\eqref{eq:delta-approximation}, replacing $\delta_{\Gamma}(\tensor{x})$ by the crack density function $\Gamma_d(d, \grad d)$.
This yields the following strain-based kinematic expression, fully compatible with the phase-field formulation
\begin{align}
	\tstrain \approx \bar{\tstrain} + \tensor{\mathcal{S}}  \zeta_{f} \Gamma_{d} (d, \grad d) \alpha .  \label{eq:strain-decompose-approx}
\end{align}

Using the strain-based fault kinematics, we formulate the interface stress by assuming that the entire $\tstress_{f}$ is contributed by the continuous deformation component of the strain~\cite{regueiro2001plane} 
\begin{align}
    \tstress_{f} = \mathbb{C}^{e}:\bar{\tstrain}. 
    \label{eq:sigmaf-bar-strain}
\end{align}
where $\mathbb{C}^{e}$ denotes the elastic stiffness tensor.
Inserting Eq.~\eqref{eq:strain-decompose-approx} into the above equation gives
\begin{align}
    \tstress_{f} = \mathbb{C}^{e}:\left[\tstrain - \tensor{\mathcal{S}} \zeta_{f} \Gamma_{d} (d, \grad d) \alpha \right] . 
    \label{eq:stress-interface}
\end{align}
To incorporate fault friction into the formulation of $\tstress_{f}$, the interface stress $\tstress_{f}$ and the slip rate $V := \dot{\zeta}_{f}$ must satisfy the following Kuhn-Tucker conditions
\begin{align}
   V \geq 0, \:\: F(\tstress_{f}) := \lvert \tau_{f} \rvert - \tau_{r}\leq 0, \:\:  V F(\tstress_{f})  = 0. 
   \label{eq:KT-yield}
\end{align}
Here, $F(\tstress_{f})$ is a yield function that ensures the resolved shear stress on the fault ($\tau_{f} := \tstress_{f} : \tensor{\mathcal{S}} $) does not exceed the frictional strength $\tau_{r}$. 
These Kuhn-Tucker conditions ensure that the fault remains in a stick state ($V = 0$) when the shear stress is below the frictional strength (\ie~$F < 0$), and slips ($V > 0$) when the shear stress equals the frictional strength (\ie~$F = 0$).
To account for pressure-dependent friction, the frictional strength $\tau_{r}$ is expressed as a function of the contact normal pressure, $p_\mathrm{N} := - \tstress_{f} : (\tensor{n} \dyadic \tensor{n})$, as
\begin{align}
    \tau_{r} &=  \mu p_\mathrm{N}, 
    \label{eq:tau-r}
\end{align}
where $\mu$ is the frictional coefficient.
This pressure-dependent frictional strength together with the yield function then introduces the coupling between the normal and shear components of the stress, which is absent in the anti-plane model but essential for the in-plane version.
Finally, by combining Eqs.~\eqref{eq:stress-interface}, \eqref{eq:KT-yield}, and \eqref{eq:tau-r}, we establish a general constitutive formulation for $\tstress_{f}$. 
The interface stress can thus be determined, provided the friction coefficient $\mu$ is specified.

\subsection{Rate- and state-dependent friction}
\label{sec:rs-friction}

We complete the formulation of the interface stress by specifying the expression for the friction coefficient. To model fault mechanics, we adopt the widely used rate- and state-dependent friction model~\cite{dieterich1979modeling,ruina1983slip}, where the friction coefficient is expressed as a function of the slip rate $V$ and a state variable $\theta$ as
\begin{align}
    \mu (V, \theta) := \mu_{0} + a\ln \left(\dfrac{V}{V_{0}}\right) + b \ln \left(\dfrac{\theta V_{0}}{D_{c}}\right).   \label{eq:rs-friction-classic}
\end{align}
Here, $\mu_{0}$ denotes the reference frictional coefficient, $a$ and $b$ are constitutive parameters quantifying the dependence on the slip rate and the state variable, respectively, $V_{0}$ is the reference slip rate, and $D_{c}$ is the characteristic slip distance.
The evolution of the state variable $\theta$ is governed by the aging law~\cite{rice1983stability}, given by 
\begin{align}
    \dot{\theta} = 1 - \dfrac{V\theta}{D_{c}} . \label{eq:aging-law}
\end{align}
This equation indicates that frictional slip reaches a steady state ($\dot{\theta} = 0$) when $V = D_{c}/\theta$. 
At steady state, the friction coefficient becomes
\begin{align}
    \mu_{ss} = \mu_{0} + (a - b) \ln\left ( \dfrac{V}{V_{0}}\right ). 
\end{align}
If $a > b$, the steady-state friction $\mu_{ss}$ increases with the slip rate, characterizing velocity-strengthening behavior. Conversely, the fault exhibits velocity-weakening friction if $a < b$.

A numerical challenge arises with the original rate- and state-dependent friction formulation~\eqref{eq:rs-friction-classic}, as it becomes ill-posed at $V = 0$. To address this, we employ the following regularized formulation proposed by Rice and Ben-Zion~\cite{rice1996slip}
\begin{align}
	\mu (V, \theta) =a \sinh^{-1}\left[\dfrac{V}{2V_{0}} \exp \left( \dfrac{\mu_{0} + b \ln\left(\theta V_{0}/D_{c} \right)}{a}\right)\right] . 
    \label{eq:rs-friction-regularize}
\end{align}
This regularized equation ensures that $\mu$ remains well-defined and non-negative, and it gradually converges to the original formulation~\eqref{eq:rs-friction-classic} as the slip rate increases.
Also notably, with this regularized formulation, the friction coefficient $\mu$ naturally vanishes at $V = 0$, meaning no fault friction in the stick state. 
As a result, the stick condition ($F < 0$) no longer needs explicit consideration. 
The yield function can thus be reformulated to directly enforce the slip condition as
\begin{align}
     F(\tstress_{f}, V) = \lvert \tau_{f} \rvert - \tau_{r} = \lvert \tau_{f} \rvert - p_\mathrm{N} \mu(V, \theta) = 0 . \label{eq:yield-slip}
\end{align}

\subsection{Quasi-dynamic fault sliding}
To incorporate inertial effects associated with fault slip, we adopt the quasi-dynamic approach, as employed in the previous anti-plane phase-field model~\cite{fei2023phase} and widely used in fault modeling studies~\cite{erickson2014efficient,thomas2014quasi,pampillon2018dynamic,abdelmeguid2019novel,erickson2020community,jiang2022community}. 
This method, originally proposed by Rice~\cite{rice1993spatio}, offers significant computational efficiency compared to fully dynamic formulations.
Specifically, instead of explicitly computing the full inertia term ($\rho\ddot{\tensor{u}}$) in Eq.~\eqref{eq:momentum-balance-pf}, the quasi-dynamic approach approximates inertial effects by incorporating a radiation damping term, $\eta V$, into the calculation of the fault shear stress, $\tau_{f}$. 
Here, $\eta$ represents the damping coefficient, typically defined as
\begin{align}
    \eta := \dfrac{G}{2 c_{s}}, \label{eq:damping-coeff}
\end{align}
where $c_{s}$ denotes the shear wave speed. 
This modification removes the inertial term, reducing the momentum balance equation to its quasi-static form as
\begin{align}
    \div \tstress(\tensor{u}, d) = \tensor{0} \quad & \text{in} \:\: \Omega . \label{eq:momentum-balance-pf-qd}
\end{align}
The yield function~\eqref{eq:yield-slip} is further modified to include the radiation damping term, as
\begin{align}
    F(\tstress_{f}, V) = \lvert \tau_{f} \rvert - p_\mathrm{N} \mu(V, \theta) - \eta V = 0 . \label{eq:yield-final}
\end{align}
In this formulation, the radiation damping term acts as an additional component of fault strength that increases with $V$. 
This term prevents infinite fault slip rates and ensures numerical stability, even in the absence of the inertial term in Eq.~\eqref{eq:momentum-balance-pf-qd}.

% SECTION 3
% ------------------------------------------------------------------------------
\section{Phase-field formulation for quasi-dynamic fault nucleation and propagation}
\label{sec:formulation}

So far, our focus has been limited to frictional slip on non-propagating faults. 
In this section, we extend the phase-field formulation to comprehensively model the nucleation and propagation of fault surfaces, as well as the growth of off-fault damage. 
Additionally, we introduce a criterion for predicting the propagation direction of faults under in-plane loading conditions.

\subsection{Phase-field equation for growing faults and damage}
We begin by referring to the governing equation formulated by Fei and Choo~\cite{fei2020phaseb} for modeling the growth of frictional shear fractures in geologic materials. 
Using the adopted crack density function~\eqref{eq:crack-density-function}, the governing equation is expressed as
\begin{align}
    -g'(d) \mathcal{H}^{+} + \dfrac{3\mathcal{G}_{II}}{8L} \left(2L^2 \div \grad d - 1 \right) = 0 \quad & \text{in} \:\: \Omega . \label{eq:pf-evolution-eq}
\end{align}
Here, $\mathcal{H}^{+}$ represents the effective crack driving force, arising from the release of stored energy and potential mechanical dissipation (\eg~frictional dissipation) during fracture propagation, and $\mathcal{G}_{II}$ denotes the critical fracture energy for shear fractures.
This equation describes fracture growth as a process in which the release of stored energy and mechanical dissipation (\ie~$-g'(d) \mathcal{H}^{+} (\tensor{u})$) is balanced by the energy required to generate new fracture surfaces (\ie~${3\mathcal{G}_{II}}/{8L} (2L^2 \div \grad d - 1)$), consistent with the classical principles of fracture mechanics.

As derived in Fei and Choo~\cite{fei2020phaseb}, the degradation function $g(d)$ associated with the adopted crack density function~\eqref{eq:crack-density-function} is expressed as
\begin{align}
    g(d) = \dfrac{(1 - d)^n}{(1 - d)^n + md(1+pd)} \quad \text{with} \:\: m = \dfrac{3\mathcal{G}_{II}}{8L} \dfrac{1}{\mathcal{H}_{t}}, \label{eq:gd}
\end{align}
where $n$ and $p$ are modeling parameters that control the degradation rate of the material as fracture/damage evolves.   
By default, we set $n=2$ and $p=1$ for numerical simulations in Section~\ref{sec:simulation}.
The term $\mathcal{H}_{t}$ represents the threshold crack driving force at the peak shear strength $\tau_{p}$ and is defined as~\cite{fei2020phaseb}
\begin{align}
    \mathcal{H}_{t} = \dfrac{1}{2G} \left[\tau_{p} - \tau_{r} \right]^2, 
\end{align}
where
\begin{align}
    \tau_{p} &= p_\mathrm{N} \mu _{p} + c,
\end{align}
with $\mu_{p}$ as the frictional coefficient for $\tau_{p}$ and $c$ as the material cohesion. 
Assuming the same frictional coefficient for both peak and residual shear strengths (\ie~$\mu_{p} = \mu$), the expression for $\mathcal{H}_{t}$ simplifies to
\begin{align}
    \mathcal{H}_{t} = \dfrac{1}{2G} c^2 . 
    \label{eq:cdf-threshold}
\end{align}
The primary reason for adopting the degradation function form in Eq.~\eqref{eq:gd} is its ability to ensure insensitivity of the peak strength and the stress-strain response to the regularization length $L$. This feature allows $L$ to function solely as a geometric parameter for phase-field regularization. Such length insensitivity is absent in most existing phase-field models. In those models, $L$ influences both the strength and softening behavior, rendering them unsuitable for simulating fracturing processes in rocks, whose strengths should remain independent of a geometric parameter.

\subsection{Crack driving force for quasi-dynamic faults}
To complete the phase-field governing equation~\eqref{eq:pf-evolution-eq}, we derive the crack driving force $\mathcal{H}^{+}$ by taking the partial derivative of the potential energy density $\psi$ (excluding fracture dissipation) with respect to $d$, say,
\begin{align}
    g'(d)\mathcal{H}^{+} = \dfrac{\pd \psi}{\pd d}. \label{eq:cdf-psi}
\end{align}
As described by Fei \etal~\cite{fei2023phase}, the potential energy density is decomposed into three components to describe quasi-dynamic fault propagation, as 
\begin{align}
    \psi = \psi^\mathrm{e} + \psi^\mathrm{f} + \psi^\mathrm{r}, \label{eq:potential-energy}
\end{align}
where $\psi^\mathrm{e}$ is the elastic strain energy density, $\psi^\mathrm{f}$ is the frictional dissipation density, and $\psi^\mathrm{r}$ represents the viscous dissipation density associated with radiation damping. 
In the context of quasi-dynamics, the viscous dissipation term approximates the energy radiated as seismic waves during earthquakes. 
The adopted decomposition of the crack driving force~\eqref{eq:potential-energy} ensures that the phase-field evolution equation~\eqref{eq:pf-evolution-eq} is intrinsically consistent with the earthquake energy budget~\cite{kanamori2000microscopic}.

To determine the specific expression for $\mathcal{H}^{+}$, we formulate each energy component in Eq.~\eqref{eq:potential-energy}. 
While these formulations for anti-plane models have been presented in Fei \etal~\cite{fei2023phase} for both stick and slip fault conditions, here we extend them to in-plane settings. 
Given that the stick condition is eliminated by employing the regularized rate- and state-dependent friction (see Section~\ref{sec:rs-friction}), we focus exclusively on the slip condition.
Given the absence of the stick condition due to the use of the regularized rate- and state-dependent friction (see Section~\ref{sec:rs-friction}), we focus solely on the slip condition. 
We express the energy densities in rate forms, as done in Fei and Choo~\cite{fei2020phaseb} and Fei \etal~\cite{fei2023phase}, to account for their incremental nonlinearity caused by evolving friction and radiation damping during fault rupture and propagation. 
Accordingly, the rate of the elastic strain energy density is formulated as
\begin{align}
    \dot{\psi}^\mathrm{e} = 
        \tstress_{m}:\dot{\tstrain} - \left[1 - g(d)\right] \tau_{m} \dot{\gamma},  \label{eq:psi-e}
\end{align}
where $\tau_{m} := \tstress_{m}:\tensor{\mathcal{S}}$ and $\gamma := 2\tstrain:\tensor{\mathcal{S}}$ denote the bulk shear stress and total shear strain in the fault slip direction, respectively.
The rate of the frictional dissipation density is
\begin{align}
    \dot{\psi}^\mathrm{f} = 
        \left[1 - g(d)\right] \tau_{r} \dot{\gamma},  \label{eq:psi-f}
\end{align}
and the rate of the viscous dissipation density is
\begin{align}
    \dot{\psi}^\mathrm{r} =
        \left[1 - g(d)\right]\eta V\dot{\gamma}. 
 \label{eq:psi-r}
\end{align}
For detailed derivations of these energy densities, see Fei and Choo~\cite{fei2020phaseb} and Fei \etal~\cite{fei2023phase}.

By integrating the energy density rates in Eqs.~\eqref{eq:psi-e}--\eqref{eq:psi-r} over time and differentiating the results with respect to $d$, we obtain the crack driving force $\mathcal{H}^{+}$. 
To ensure no damage growth before the intact material reaches the peak strength $\tau_{p}$, we set $\mathcal{H}^{+} = \mathcal{H}_{t}$ at $d = 0$. 
This results in the following expression for $\mathcal{H}^{+}$
\begin{align}
    \mathcal{H}^{+} = \left \{ 
    \begin{array}{ll}
        \mathcal{H}_{t} & \text{if}\:\:\text{intact},  \\
        \mathcal{H}_{t} + \mathcal{H}_\mathrm{slip}  & \text{if}\:\:\text{fractured}, 
    \end{array}
    \right . \quad \text{with} \:\: \mathcal{H}_\mathrm{slip} := \int_{\gamma_{p}}^{\gamma} (\tau_{m} - \tau_{r} - \eta V ) \: \dd \gamma .  \label{eq:crack-driving-force}
\end{align}
Here, $\mathcal{H}_\mathrm{slip}$ represents the crack driving force arising from quasi-dynamic frictional slip, and $\gamma_{p}$ denotes the total shear strain in the slip direction when the material strength is reached.  

\subsection{Fault propagation direction in an in-plane setting}

The remaining task is to determine the directional vectors, $\tensor{n}$ and $\tensor{m}$, for potential fault slip surfaces. 
Following the approach used in the previous anti-plane model, we hypothesize that fault propagation occurs in the direction where the crack driving force reaches its maximum at the onset of fault growth. 
This assumption is consistent with common practices in other phase-field formulations~\cite{bryant2018mixed,fei2020phaseb,fei2021double}.

Since the increment of the crack driving force is proportional to $(\tau_{m} - \tau_{r} - \eta V)$, as shown in Eq.~\eqref{eq:crack-driving-force}, we model fault propagation to maximize this term. 
By assuming steady-state conditions and $V = V_{0}$ in the intact region ($d = 0$), the radiation damping term is negligible, reducing the problem to maximizing $(\tau_{m} - p_\mathrm{N} \mu_{0})$.
In the anti-plane model, this criterion simplifies further to maximizing $\tau_{m}$, as the contact normal stress remains a prescribed constant under anti-plane loading. 
However, for in-plane modeling, the contact normal stress evolves with applied loading, making the problem more complex. 
Thus, the goal in the in-plane case is to identify the fault plane where $(\tau_{m} - p_\mathrm{N} \mu_{0})$ attains its maximum.
This criterion is identical to that in the phase-field model for frictional shear fractures under quasi-static conditions~\cite{fei2020phaseb}. 
For brevity, we refer to the derivation procedure in Fei and Choo~\cite{fei2020phaseb} and present the final expressions for the directional vectors
\begin{align}
    \tensor{n} = \tensor{a}_{1} \sin \left[\dfrac{\pi- 2\arctan(\mu_{0})}{4} \right] + (\tensor{a}_{2} \times \tensor{a}_{1}) \cos \left[\dfrac{\pi- 2\arctan(\mu_{0})}{4} \right] , 
\end{align}
and 
\begin{align}
    \tensor{m} = \tensor{a}_{2} \times \tensor{n}, 
\end{align}
where $\tensor{a}_{1}$ is the direction of the major principal stress in compression, and $\tensor{a}_{2}$ is the out-of-plane direction.
Notably, compared to the anti-plane model (see Eq. (42) in Fei \etal~\cite{fei2023phase}), the normal vector in the in-plane model depends not only on local stress conditions but also on the reference friction coefficient, $\mu_{0}$.

\subsection{Summary of equations}
The proposed phase-field model for a quasi-dynamic, in-plane condition consists of two governing equations. 
The first is the momentum balance equation
\begin{align}
    \div \tstress(\tensor{u}, d) = \tensor{0} \quad  \text{in} \:\: \Omega , 
    \label{eq:momentum-balance-final} 
\end{align}
with the boundary conditions
\begin{align}
    \tstress \cdot \tensor{v} = \hat{\tensor{t}} \quad &\text{on} \:\: \pd_{t} \Omega, \\ 
	\tensor{u} = \hat{\tensor{u}} \quad & \text{on} \:\: \pd_{u} \Omega. 
\end{align}
The second is the phase-field evolution equation
\begin{align}
    -g'(d) \mathcal{H}^{+}+ \dfrac{3\mathcal{G}_{II}}{8L} \left(2L^2 \div \grad d - 1 \right) = 0 \quad & \text{in} \:\: \Omega , 
    \label{eq:pf-evolution-final}
\end{align}
with the boundary condition
\begin{align}
    \grad d = \tensor{0} \quad & \text{on} \:\: \pd \Omega.
\end{align}
The evaluations of $\tstress$ and $\mathcal{H}^{+}$ depend on the state of the material, specifically whether it is intact ($d = 0$) or fractured ($d > 0$). 
Their respective expressions are
\begin{align}
    \left .
    \begin{array}{r}
         \dot{\tstress} = \mathbb{C}^{e}:\dot{\tstrain } \vspace{1em} \\
         \mathcal{H}^{+} = \mathcal{H}_{t}
    \end{array}
    \right \} \text{if}\:\:\text{intact,}
\end{align}
and 
\begin{align}
    \left .
    \begin{array}{r}
         \dot{\tstress} =  \mathbb{C}^{e}:\dot{\tstrain} - [1 - g(d)]  \dot{\zeta_{f} }\Gamma_{d} (d, \grad d) \alpha \mathbb{C}^{e}:\tensor{\mathcal{S}} \vspace{1em} \\
         \mathcal{H}^{+} = \mathcal{H}_{t} + \int_{\gamma_{p}}^{\gamma} (\tau_{m} - \tau_{r} - \eta V ) \: \dd \gamma 
    \end{array}
    \right \} \text{if}\:\:\text{fractured.}
\end{align}
Here, $\tstress$ is expressed in a rate form to account for the nonlinear evolution and history dependence of the stress tensor. 
For fractured material, the rate of slip magnitude $\dot{\zeta}_{f}$ is calculated by satisfying the yield function in Eq.~\eqref{eq:yield-final}.

The proposed formulation can be solved numerically using the standard finite element method.
Details of the discretization algorithms to numerically solve Eqs.~\eqref{eq:momentum-balance-final} and \eqref{eq:pf-evolution-final} are presented in~\ref{sec:appendix}.

% SECTION 4
% ------------------------------------------------------------------------------
\section{Verification}
\label{sec:verification}

In this section, we verify the proposed phase-field formulation for modeling quasi-dynamic fault rupture. 
To achieve this, we compare the phase-field simulation results for in-plane fault rupture with those obtained from a well-validated discontinuous approach. 
The material parameters used in the verification example are summarized in Table~\ref{tab:material-parameters}, with specific values adopted from the SCEC/USGS verification example TPV102~\cite{harris2009scec,harris2018suite}.
Based on the mass density and shear wave speed provided in the table, the shear modulus is calculated as $G = \rho c_{s}^2 = 32.038$ GPa. 
This calculation yields a radiation damping coefficient of $4.624$ MPa$\cdot$s/m, determined using Eq.~\eqref{eq:damping-coeff}.
Furthermore, as shown in the table, the parameters satisfy $a < b$, indicating that the simulated fault exhibits velocity-weakening friction. 

\begin{table}[htbp]
    \centering
    \begin{tabular}{l|c|c|c}
    \hline 
      \textbf{Parameter}   & \textbf{Symbol} & \textbf{Unit} & \textbf{Value}  \\
    \hline
      Mass density   & $\rho$ & kg/m$^3$ & 2670 \\ 
      Shear wave speed & $c_{s}$ & km/s & 3.464 \\
      Bulk modulus& $K$ & GPa & 69.416 \\ 
      Parameter for rate-dependence& $a$ & - & 0.008 \\ 
      Parameter for state-dependence & $b$ & - & 0.012 \\ 
      Reference slip rate & $V_{0}$ & m/s & $10^{-6}$ \\
      Characteristic slip distance & $D_{c}$ & m & 0.02 \\
      Default reference frictional coefficient & $\mu_{0}$ & - & 0.6 \\ 
      \hline
    \end{tabular}
    \caption{Material parameters used in all simulations.}
    \label{tab:material-parameters}
\end{table}

\begin{figure}[htbp]
    \centering
    \includegraphics[width=\textwidth]{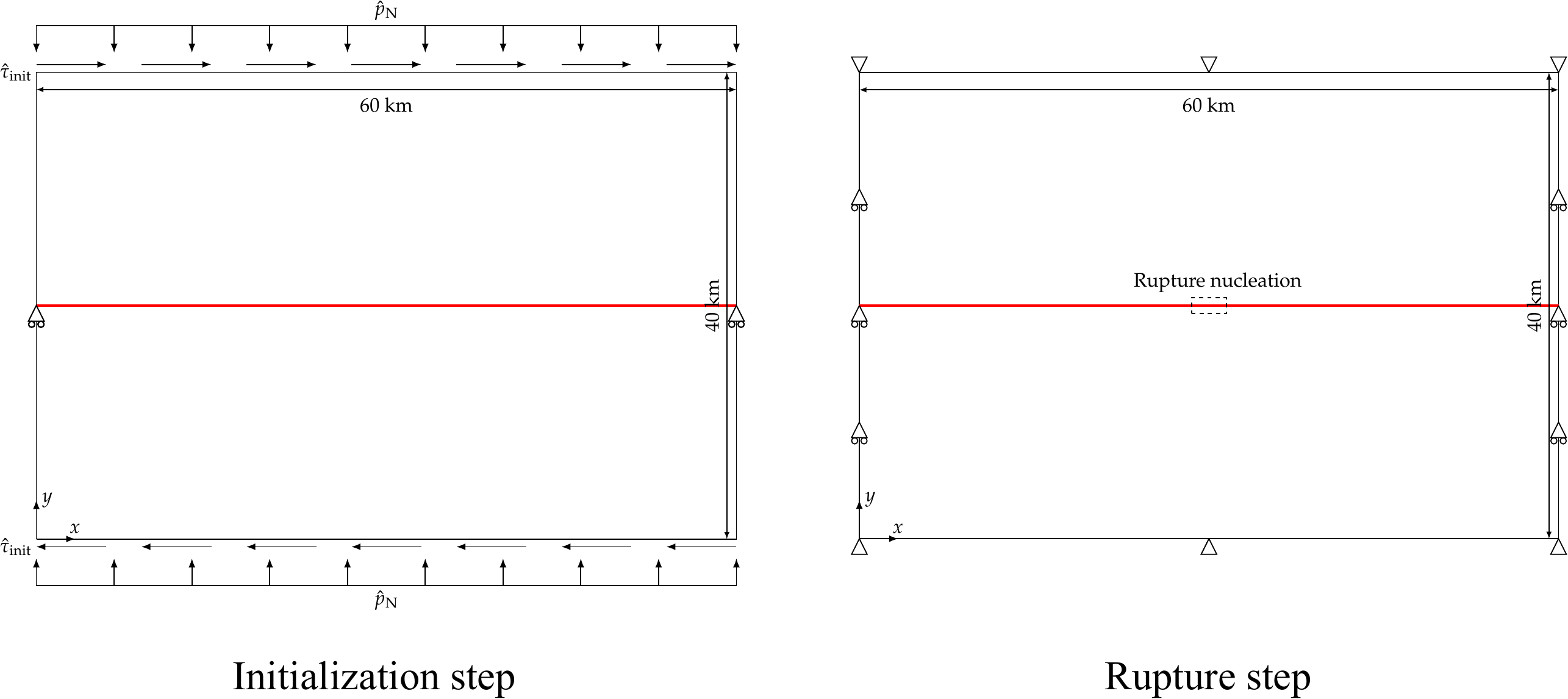}
    \caption{Quasi-dynamic rupture of a non-growing fault: problem setup.}
    \label{fig:rupture-setup}
\end{figure}

The problem setup is adapted from the benchmark example for the anti-plane model~\cite{fei2023phase}, with modifications to the boundary conditions to suit the in-plane configuration. 
Figure~\ref{fig:rupture-setup} depicts the domain geometry and boundary conditions of the problem. 
The domain is a rectangular region, 60 km wide and 40 km high, containing a horizontal fault located at $y = 20$ km, spanning the full width of the domain.  
In the phase-field simulation, the horizontal fault is modeled as a diffuse damage field, initialized by prescribing a high crack driving force value at the expected fault location, following the methodology of Borden \etal~\cite{borden2012phase}.
Additionally, we set $p=20$ in the degradation function~\eqref{eq:gd}, as a higher $p$ improves the accuracy of the phase-field results and facilitates convergence studies~\cite{geelen2019phase}.
% To ensure accuracy while avoiding unnecessary computational expense, we locally refine the region where the fault is seated. 
The simulation consists of two distinct steps, each characterized by different boundary conditions: an initialization step and a rupture step. 
Details of the purpose and boundary conditions applied in each step are provided in the following paragraphs.

The simulation begins with an initialization step to establish the normal and shear stresses within the implicitly represented phase-field interface. 
This is accomplished by applying traction boundary conditions on the top and bottom surfaces, as shown in Fig.~\ref{fig:rupture-setup}, and solving the momentum balance equation to update the stress distribution. 
To ensure domain stability, roller constraints are imposed at the centers of the lateral boundaries.
The normal traction, $\hat{p}_\mathrm{N}$, is set to 50 MPa, resulting in an initial contact normal stress on the fault of $p_\mathrm{N, init} = 50$ MPa.
Assuming the fault is in a steady state (\ie~$\dot{\theta} = 0$) with an initial slip rate of $V_\text{init} = 10^{-9}$ m/s, the initial frictional coefficient is calculated as
\begin{align}
    \mu_\mathrm{init} = a \sinh^{-1}\left[\dfrac{V_\text{init}}{2V_{0}} \exp \left( \dfrac{\mu_{0} + b \ln\left( V_{0}/V_\text{init} \right)}{a}\right)\right] \approx 0.628 . 
\end{align}
Substituting this frictional coefficient along with $p_\mathrm{N, init} = 50$ MPa and $V_\text{init} = 10^{-9}$ m/s into the yield function~\eqref{eq:yield-slip} gives an initial shear stress of $\tau_{f, \text{init}} = 31.382$ MPa. 
To achieve this expected shear stress on the fault, the applied shear traction, $\hat{\tau}_\text{init}$, is set to 31.382 MPa.

Following the initialization step, the simulation transitions to the rupture step, during which a 6-km-wide rupture nucleation zone is introduced at the center of the fault. 
To trigger fault rupture, the reference friction coefficient $\mu_{0}$ is reduced within the nucleation zone according to the following formulation
\begin{align}
  \mu_{0} (x,t) = \mu_{0, \text{init}} - (\mu_{0, \text{init}} - \mu_{0, \min} )  F(x) H(t),  
\end{align}
where $F(x)$ and $H(t)$ are smooth functions controlling the spatial and temporal evolution of $\mu_{0}$, respectively. 
These functions are defined as
\begin{align}
  F(x) =  \left \{
    \begin{array}{ll}
      \exp \left[
      \dfrac{(x - 30)^2}{(x - 30)^2 - 9} \right] & \text{if} \:\: 27\; \text{km} < x < 33 \; \text{km} ,  \\ 
      0 & \text{otherwise} , 
    \end{array}
  \right. 
\end{align}
and 
\begin{align}
  H(t) = \left \{
    \begin{array}{ll}
     0 & \text{if} \:\: t = 0 \; \text{s} ,\\
      \exp\left[\dfrac{(t - 1)^2}{t(t - 2)} \right] & \text{if} \:\: 0 \; \text{s} < t < 1 \; \text{s} , \\ 
      1 & \text{if} \:\: t > 1 \; \text{s} . 
    \end{array}
  \right. 
\end{align}
During the rupture step, the top and bottom boundaries are constrained to preserve the initialized normal and shear stresses on the fault, while the lateral boundaries are fixed in the $y$ direction.

For verification, we compare the phase-field simulation results with reference solutions generated by the hybrid finite element and spectral boundary integral (FEBE) method, which has been extensively verified in the literature~\cite{abdelmeguid2019novel,ma2019hybrid}. 
The FEBE method typically employs a pre-defined, finite-element-discretized fault zone, representing the fault as a sharp discontinuity along inter-element boundaries.
To improve computational efficiency, the FEBE method incorporates a spectral boundary integral equation to truncate the domain, assuming the truncated region to be homogeneous and linear elastic. 
However, in this verification example, the spectral boundary integral equation is deactivated, and the entire domain is fully discretized using finite elements.

\begin{figure}[htbp]
    \centering
    \subfloat[]{\includegraphics[width=0.48\textwidth]{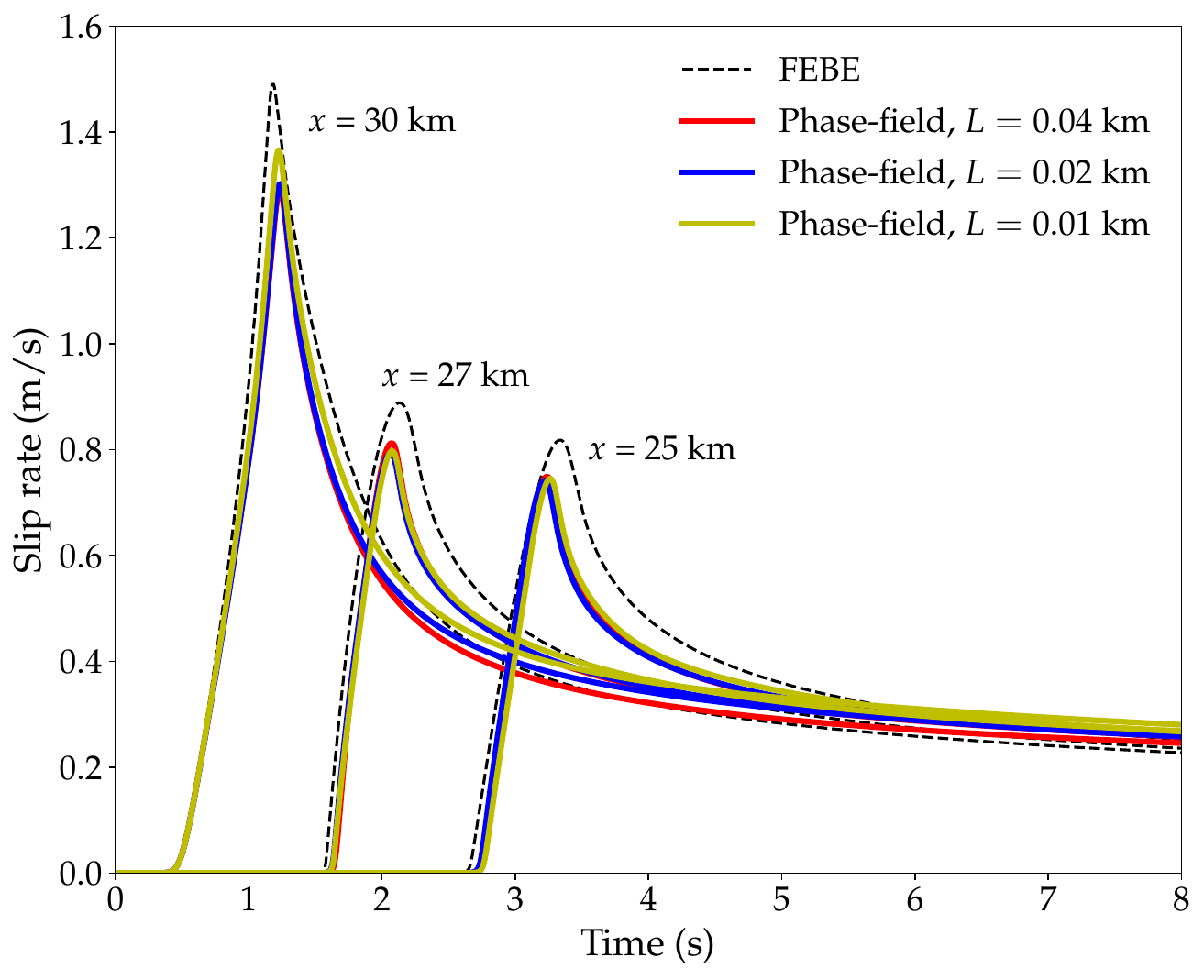}}
    \hspace{0.02\textwidth}
    \subfloat[]{\includegraphics[width=0.48\textwidth]{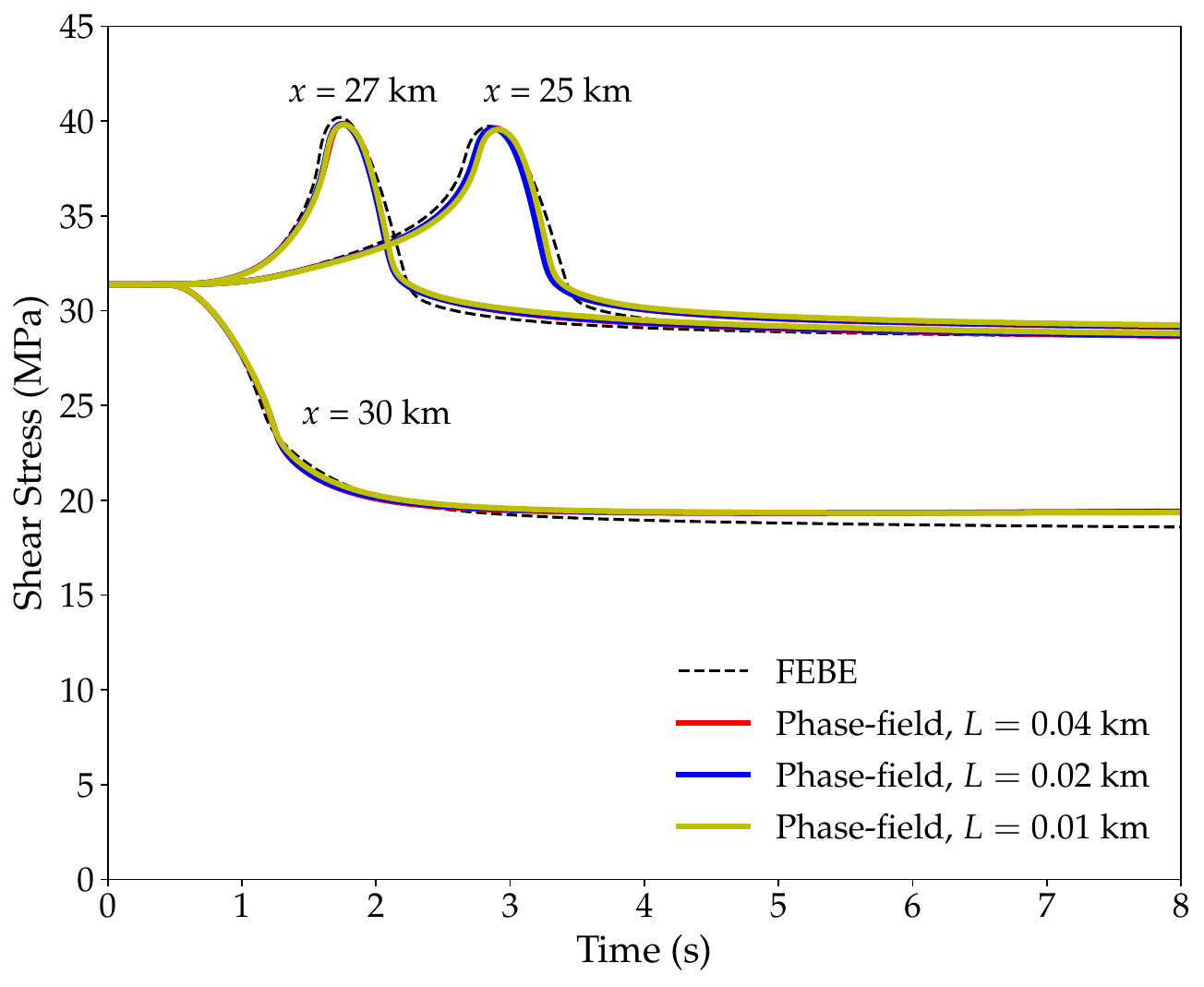}}
    \caption{Quasi-dynamic rupture of a non-growing fault: phase-field results for varying regularization lengths $L$ with fixed $L/h = 10$.}
    \label{fig:length-comparison}
\end{figure}

Figure~\ref{fig:length-comparison} illustrates the evolution of slip rates and shear stresses from the phase-field simulations, conducted with three different regularization lengths while keeping the time step size fixed at $\Delta t = 0.004$ s and maintaining the element size $h$ such that $L/h = 10$. 
Slip rates and shear stresses are measured at three fault locations: $x = 30$ km (center of the nucleation zone), $x = 27$ km (edge of the nucleation zone), and $x = 25$ km (outside the nucleation zone). 
The results show that the phase-field method effectively captures the essential kinematics observed in the FEBE solutions. 
Specifically, both approaches indicate that the slip rates at the three locations follow a trend of initial increase, subsequent decrease, and eventual stabilization at a residual state as the fault rupture propagates beyond the nucleation zone.
A similar trend is observed in the shear stress plots at $x = 25$ km and 27 km for both phase-field and FEBE simulations.  
However, at $x = 30$ km, the shear stress plot shows only a stress drop, corresponding to the reduction of $\mu_{0}$ within the nucleation zone. 
Overall, the phase-field results show good agreement with the FEBE solutions, apart from a consistent discrepancy that the phase-field model predicts a slightly slower peak slip rate. 
Notably, this discrepancy diminishes as $L$ decreases, indicating the convergence of the phase-field formulation to the classical discontinuous approach with smaller regularization lengths. 

\begin{figure}[htbp]
    \centering
    \subfloat[]{\includegraphics[width=0.48\textwidth]{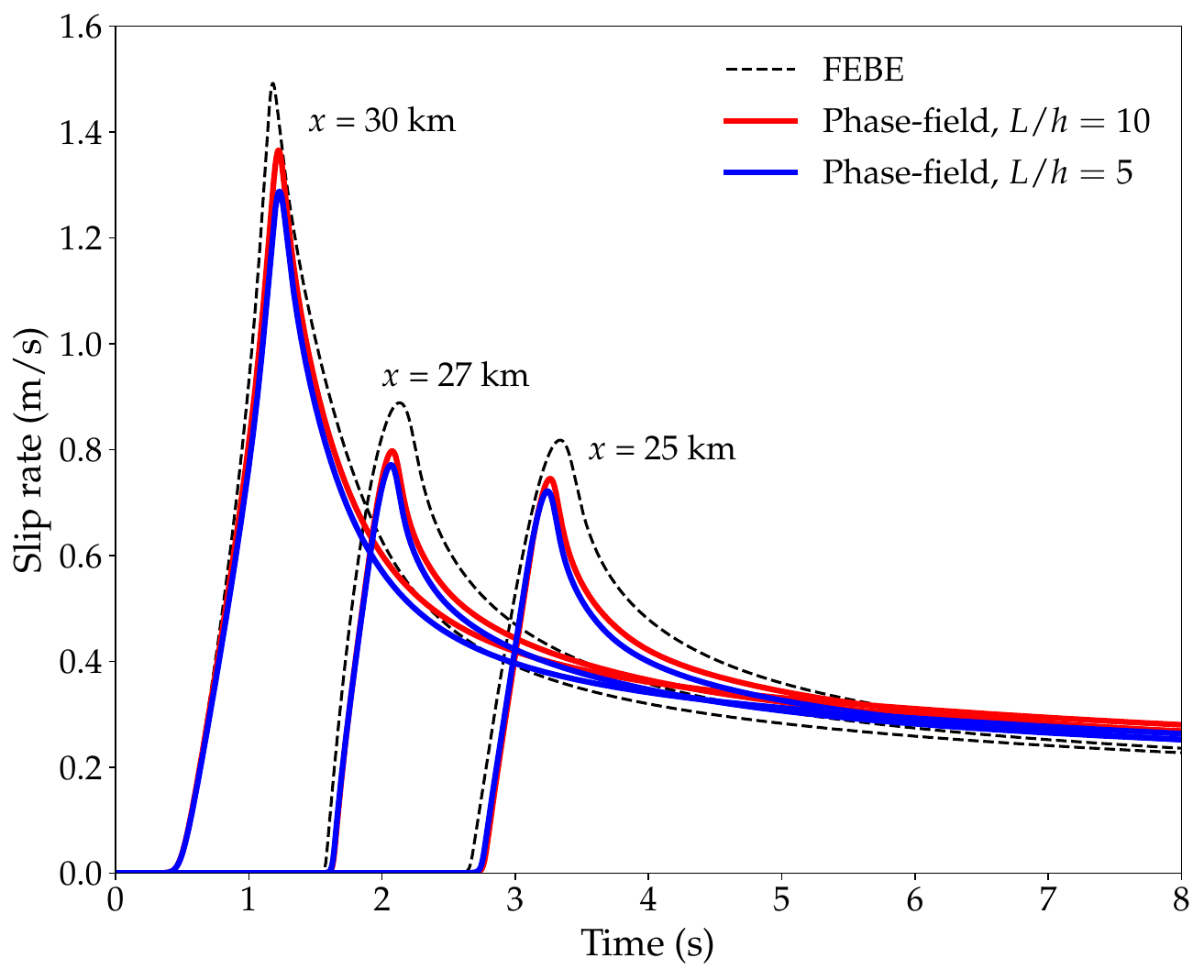}}
    \hspace{0.02\textwidth}
    \subfloat[]{\includegraphics[width=0.48\textwidth]{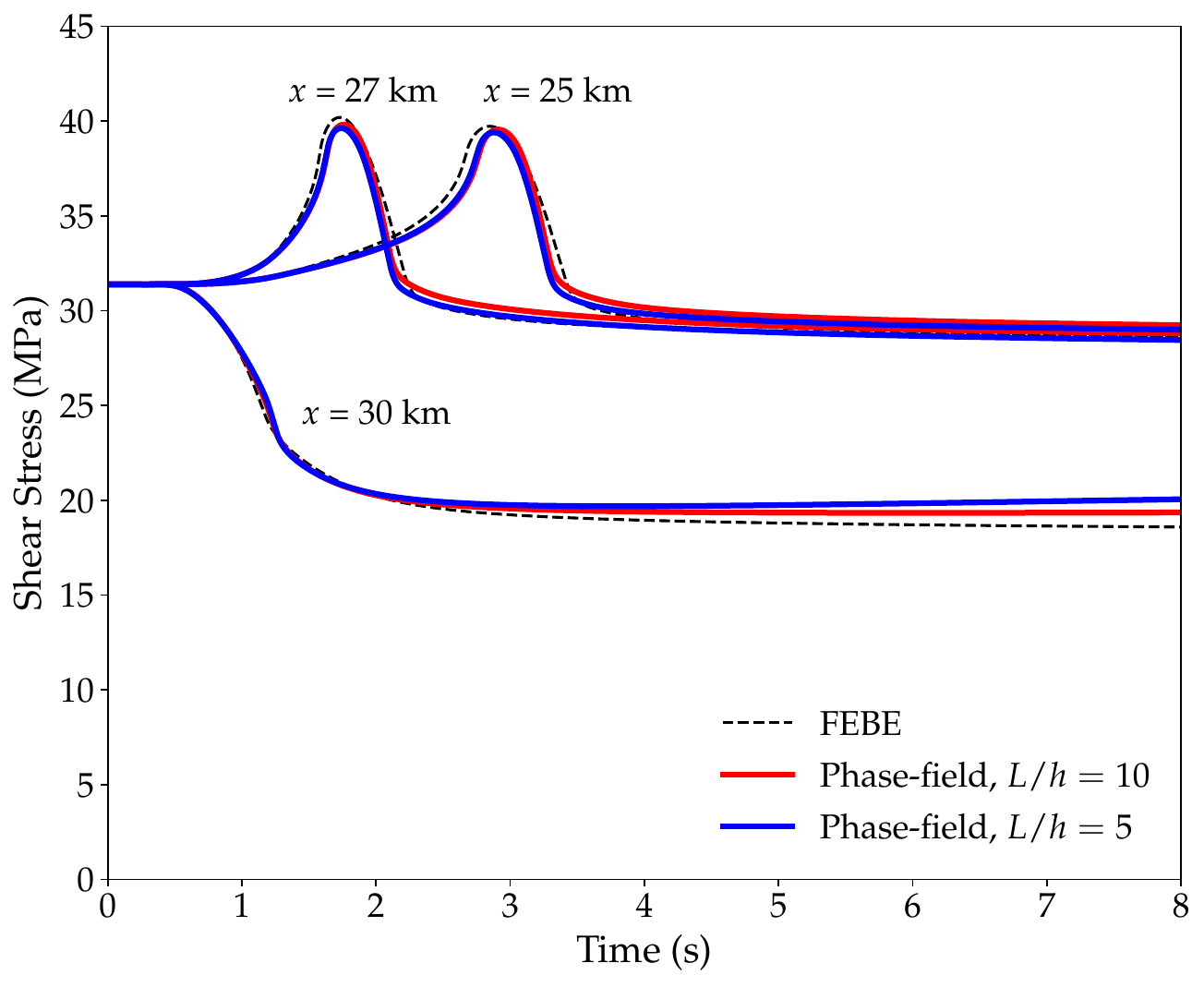}}
    \caption{Quasi-dynamic rupture of a non-growing fault: phase-field results for varying element sizes $h$ with fixed $L = 0.01$ km.}
    \label{fig:mesh-comparison}
\end{figure}

\begin{figure}[htbp]
    \centering
    \subfloat[]{\includegraphics[width=0.48\textwidth]{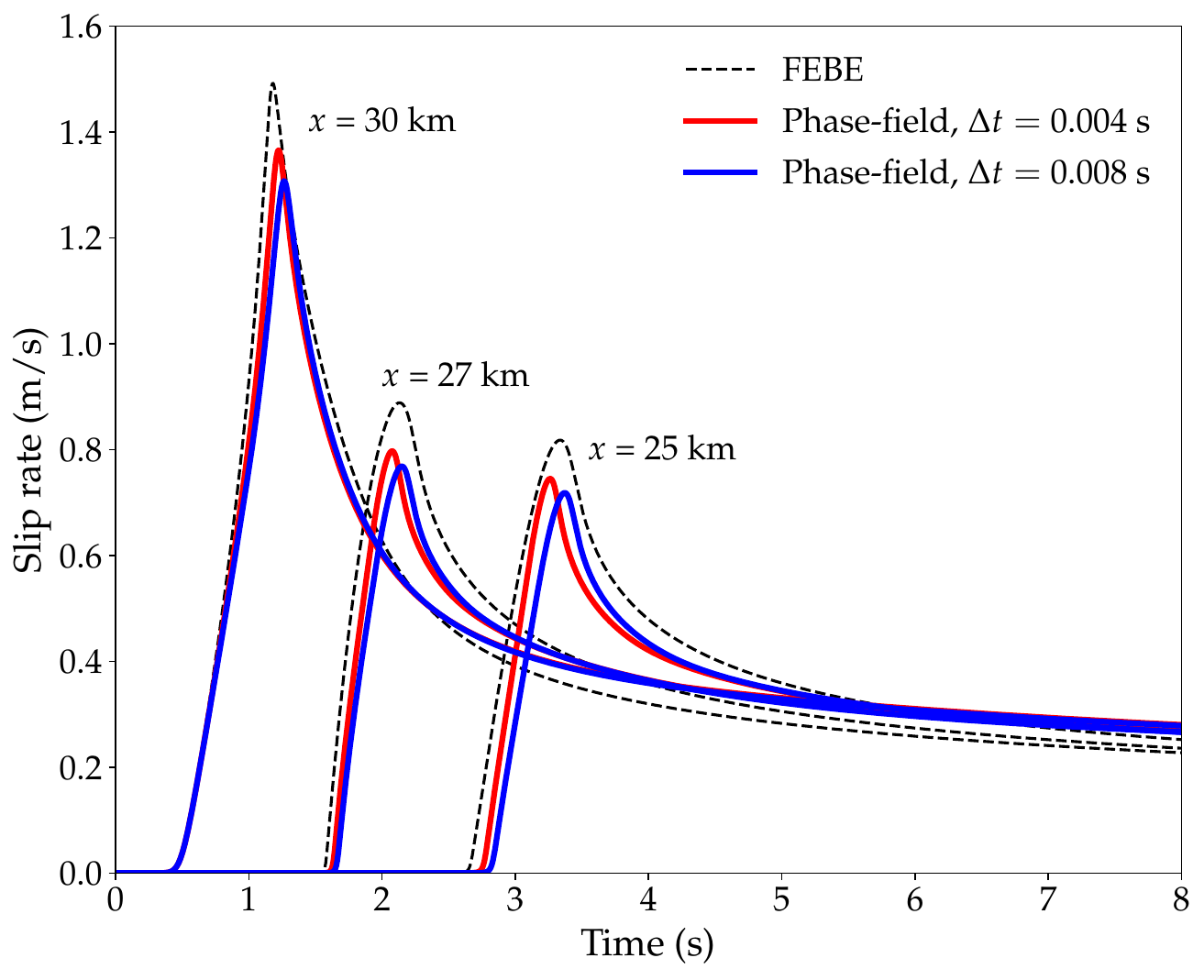}}
    \hspace{0.02\textwidth}
    \subfloat[]{\includegraphics[width=0.48\textwidth]{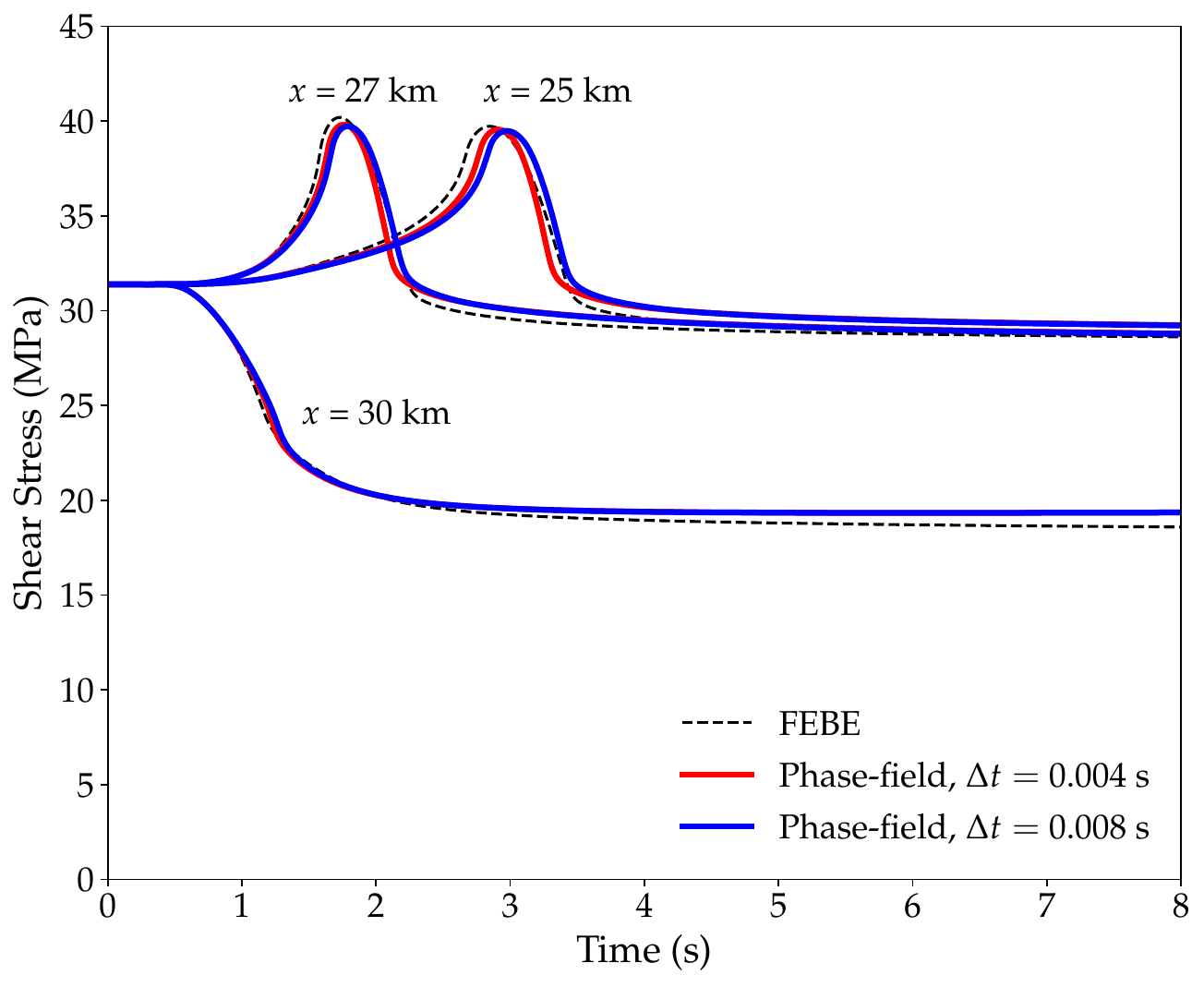}}
    \caption{Quasi-dynamic rupture of a non-growing fault: phase-field results for varying time step sizes $\Delta t$.}
    \label{fig:time-comparison}
\end{figure}

Next, we examine the numerical convergence of the phase-field model with respect to spatial and temporal discretization.
Figure~\ref{fig:mesh-comparison} presents the slip rates and shear stresses for two cases with different element sizes, specifically \ie~$L/h=5$ and 10. 
As shown, the results for both slip rate and shear stress exhibit minimal variation between the two cases.
Notably, the peak slip rate at each measurement location increases slightly and aligns more closely with the reference solution as $h$ decreases, demonstrating the convergence of the phase-field formulation with mesh refinement.
This mesh convergence is a characteristic feature of phase-field methods, attributable to their nonlocal regularized formulation.
To assess the influence of temporal discretization, we re-simulate the problem using a larger time step size of $\Delta t = 0.008$ s. 
Figure~\ref{fig:time-comparison} shows that both slip rates and shear stresses from the phase-field model asymptotically approach the FEBE solutions as the time step size decreases. 
This observation confirms the convergence of the phase-field solutions with temporal refinement.

In summary, the proposed phase-field formulation has been successfully verified for modeling quasi-dynamic fault rupture under in-plane loading conditions. 
The phase-field solutions demonstrated robust numerical convergence with respect to the regularization length, as well as spatial and temporal discretizations.
Accordingly, we conclude that the phase-field model can effectively resolve the quasi-dynamic rate- and state-dependent fault rupture problem, provided that the numerical parameters (\eg~$L$, $h$, $\Delta t$) are appropriately selected.

% SECTION 5
% ------------------------------------------------------------------------------
\section{Simulation of fault nucleation and propagation}
\label{sec:simulation}

Having verified the phase-field model, we apply it to simulate the nucleation and propagation of fault slip surfaces (or shear bands) in a series of in-plane examples with increasing structural and material complexities. Alongside these simulations, we provide interpretations of the resulting fault patterns and mechanical responses, aiming to validate further the effectiveness of the phase-field method for modeling quasi-dynamic faulting processes. The material parameters associated with fault rupture remain consistent with those listed in Table~\ref{tab:material-parameters}.

\subsection{Nucleation and propagation of faults from a weak zone}
\begin{figure}
    \centering
    \includegraphics[width=\textwidth]{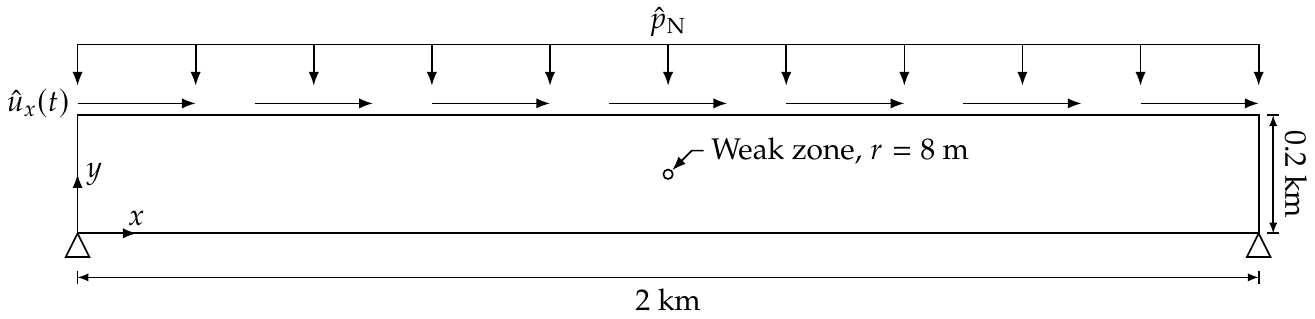}
    \caption{Nucleation and propagation of faults from a weak zone: problem setup.}
    \label{fig:fault-growth-setup}
\end{figure}

We begin with a simple case that describes fault nucleation and propagation originating from a local weak zone. 
The problem domain and boundary conditions are detailed in Fig.~\ref{fig:fault-growth-setup}. 
This setup is inspired by experiments on fault gouge~\cite{mair1999friction}, where a layer of geologic material is subjected to an in-plane continuous shear displacement, $\hat{u}_{x}(t)$, and a uniform normal pressure, $\hat{p}_\mathrm{N}$, applied at the top boundary, while the bottom boundary remains fixed in all directions. 
In this example, we prescribe $\hat{u}_{x}(t) = 0.002$ m/s and a time step size $\Delta t = 0.05$ s.
For modeling fault growth, the cohesion strength is set to $c = 10$ MPa, and the critical fracture energy $\mathcal{G}_{II} = 0.2$ MJ/m$^2$. 
To initiate fault nucleation, the domain contains a circular weak zone at its center, characterized by a reduced critical fracture energy ($\mathcal{G}_{II} = 0.1$ MJ/m$^2$) at the center, as illustrated in Fig.~\ref{fig:fault-growth-setup}. 
For each simulation, the regularization length is set to $L = 0.004$ km, and the domain is globally refined to maintain a discretization level of $L/h = 5$. 

\begin{figure}
    \centering
    \subfloat[]{\includegraphics[width=0.5\textwidth]{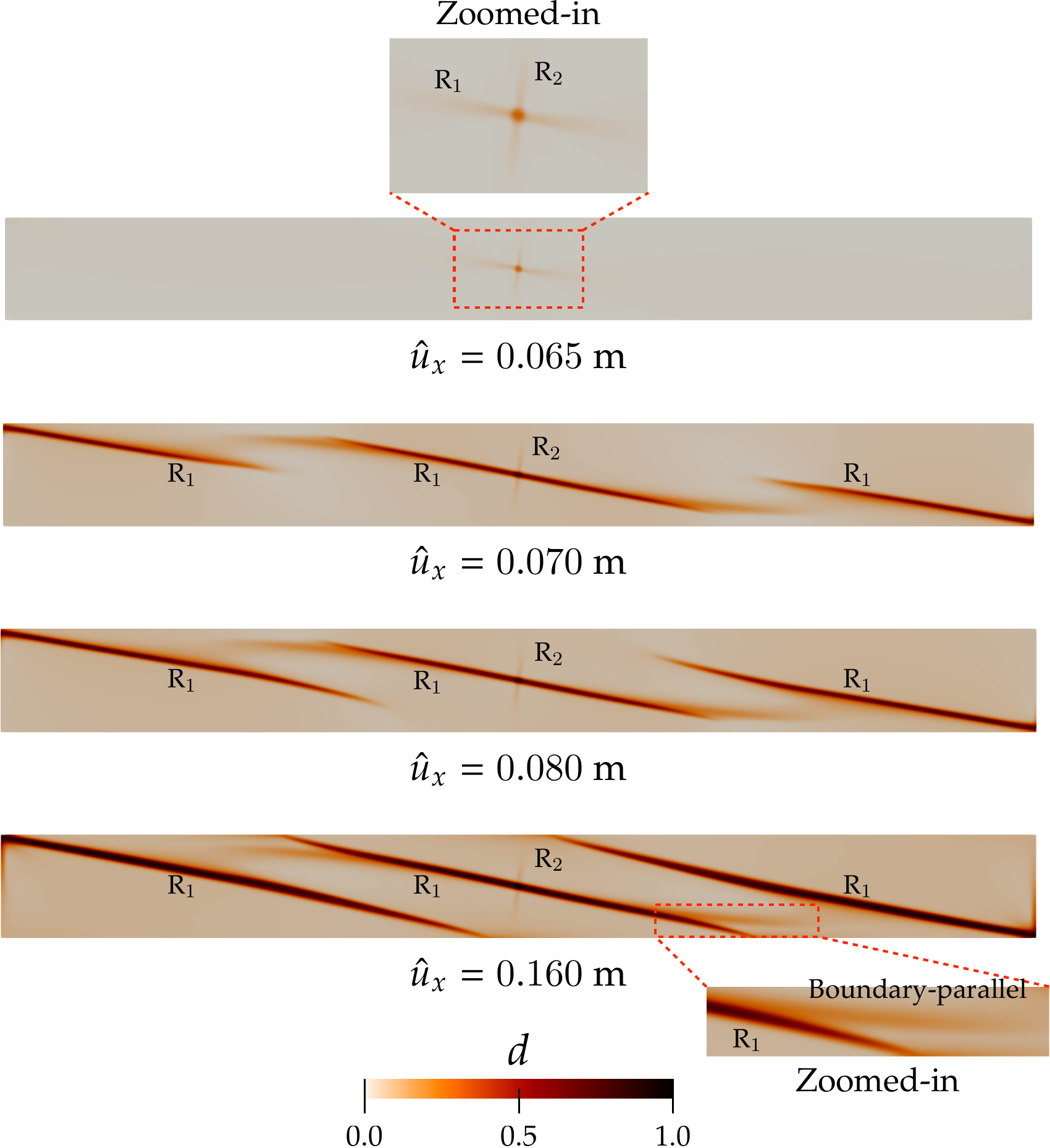}}
    \hspace{0.03\textwidth}
    \subfloat[]{\includegraphics[width=0.46\textwidth]{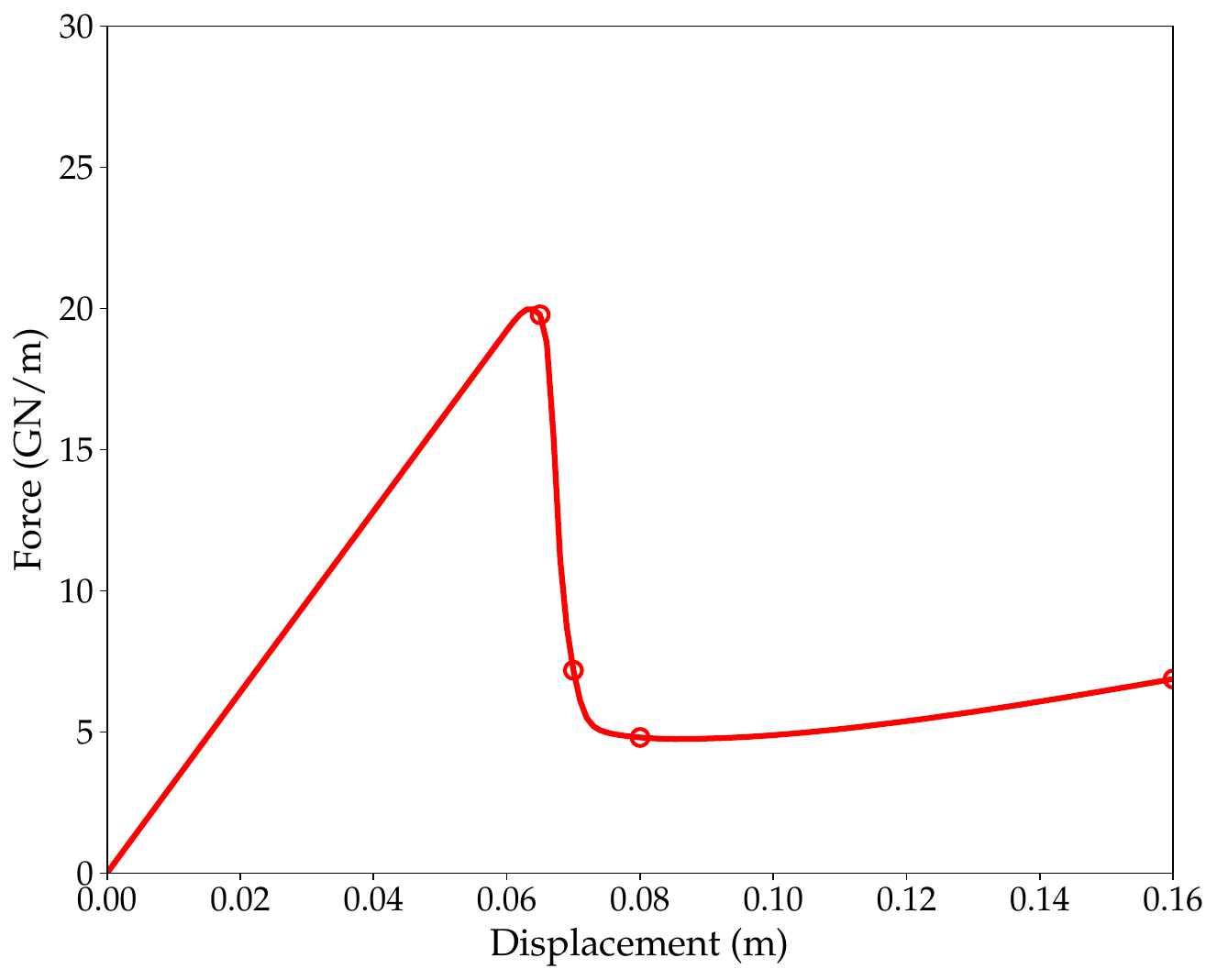}}
    \caption{Nucleation and propagation of faults from a weak zone: (a) phase-field evolution and (b) force-displacement curve for the case with $\hat{p}_\mathrm{N}=20$ MPa.}
    \label{fig:fault-growth-damage-20MPa}
\end{figure}

Figure~\ref{fig:fault-growth-damage-20MPa} illustrates the development of fault slip surfaces for the case with $\hat{p}_\mathrm{N} = 20$ MPa, along with the evolution of the applied shear force on the top boundary. 
Each stage of fault growth is annotated on the force-displacement curve to aid in interpreting the faulting process.
The results show that the nucleation of conjugate Riedel shear bands (R$_{1}$ and R$_{2}$) begins at the weak zone when the shear force reaches its peak value. 
As shear displacement increases, the system undergoes a stress drop, and the propagation of the primary Riedel shear band (R$_{1}$) becomes dominant over the secondary one (R$_{2}$). 
Additionally, continued shearing induces stress concentration at the left-top and right-bottom corners, resulting in the formation of additional shear bands in these regions.
These corner-initiated shear bands, combined with the central R$_{1}$ band, eventually form an en \'{e}chelon, overstepping array of slip surfaces---a characteristic feature of strike-slip fault systems.
At the final stage of loading, shear bands oriented more parallel to the boundaries emerge near the top and bottom surfaces, where branching of the primary Riedel shear band occurs. 
This observation aligns with the numerical findings of Ma and Elbanna~\cite{ma2018strain}. These boundary-parallel shear bands continue to grow with ongoing shearing and tend to coalesce with the R$_{1}$ slip surfaces originating from the corners.

\begin{figure}[h!]
    \centering
    \includegraphics[width=0.75\textwidth]{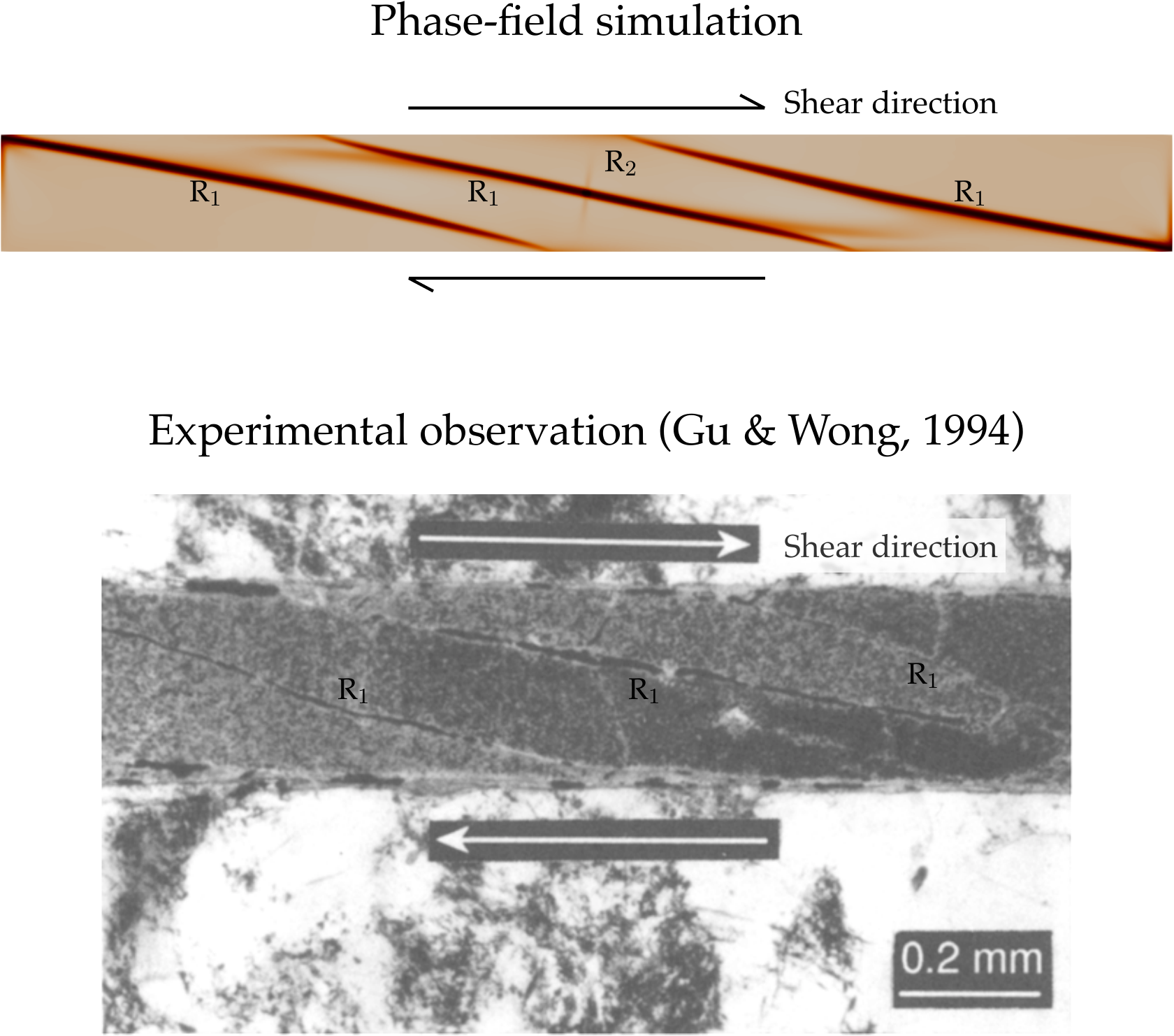}
    \caption{Nucleation and propagation of faults from a weak zone: comparison of the phase-field simulated slip surfaces (shear bands) with experimental observations.}
    \label{fig:fault-growth-damage-compare}
\end{figure}

Next, we compare the phase-field results with experimental observations to validate the model's ability to accurately capture typical slip surface patterns under in-plane shear loads. Figure~\ref{fig:fault-growth-damage-compare} contrasts the phase-field results with experimentally observed shear localization in gouge materials subjected to continuous shear~\cite{gu1994development}.
Both the simulation and experimental results show that primary Riedel shear bands (R$_{1}$) dominate the gouge material and propagate across the entire layer. Most of these R$_{1}$ shear bands exhibit curved propagation paths, with higher inclinations relative to the shear direction at the domain center, transitioning to orientations more parallel to the shear direction near the top and bottom boundaries.
This agreement between the simulation and experimental results highlights the proposed method's capability to reliably model the nucleation and propagation of fault slip surfaces under in-plane shear loads.
It is worth noting that despite the scale difference between the simulation and experiment, the phase-field model is expected to produce qualitatively similar results when appropriately scaled material parameters (\eg~$\mathcal{G}_{c}$) are used.

\begin{figure}[h!]
    \centering
    \subfloat[]{\includegraphics[width=0.52\textwidth]{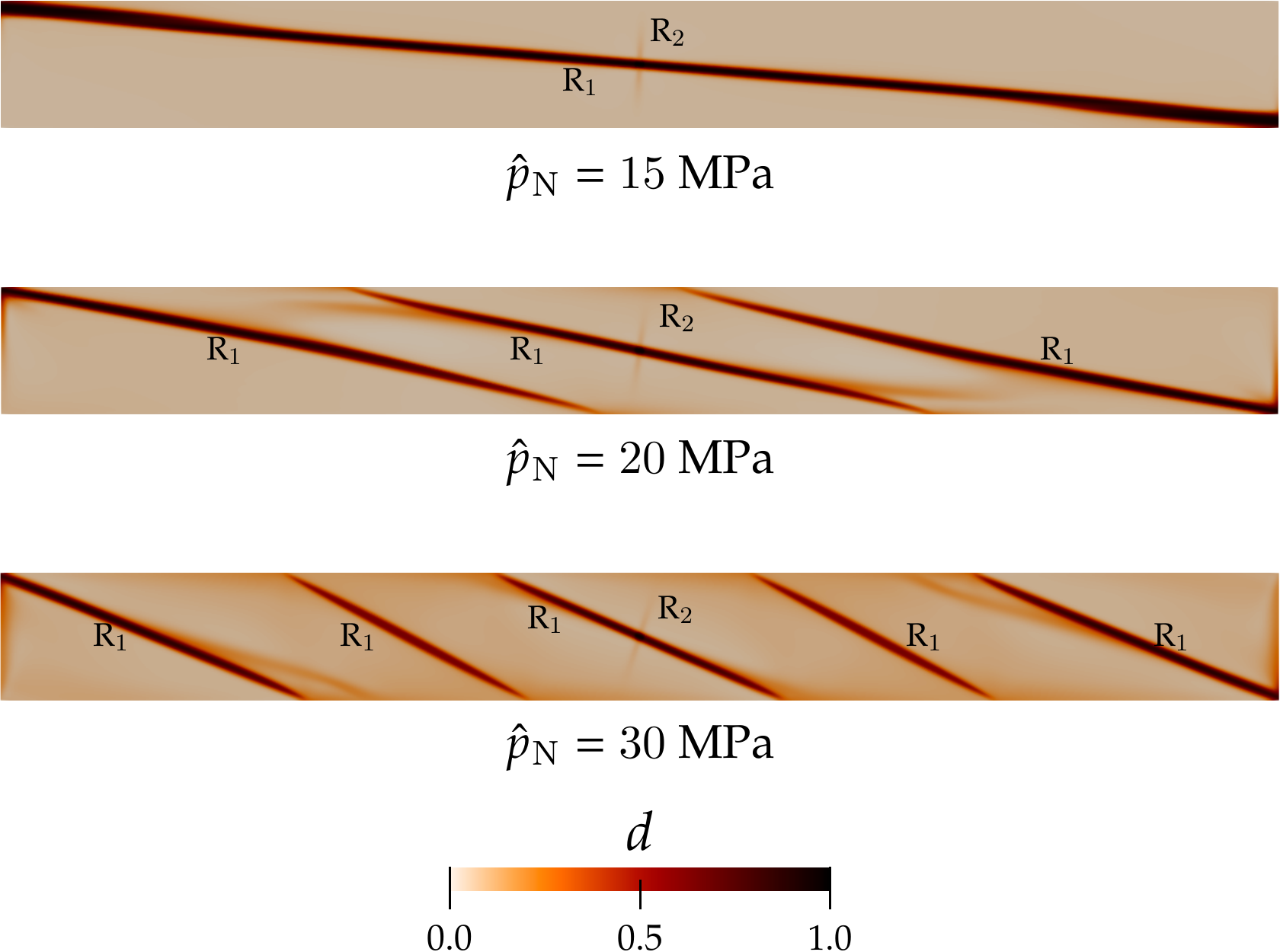} \label{fig:fault-growth-stressCompare-damage}}
    \hspace{0.02\textwidth}
    \subfloat[]{\includegraphics[width=0.43\textwidth]{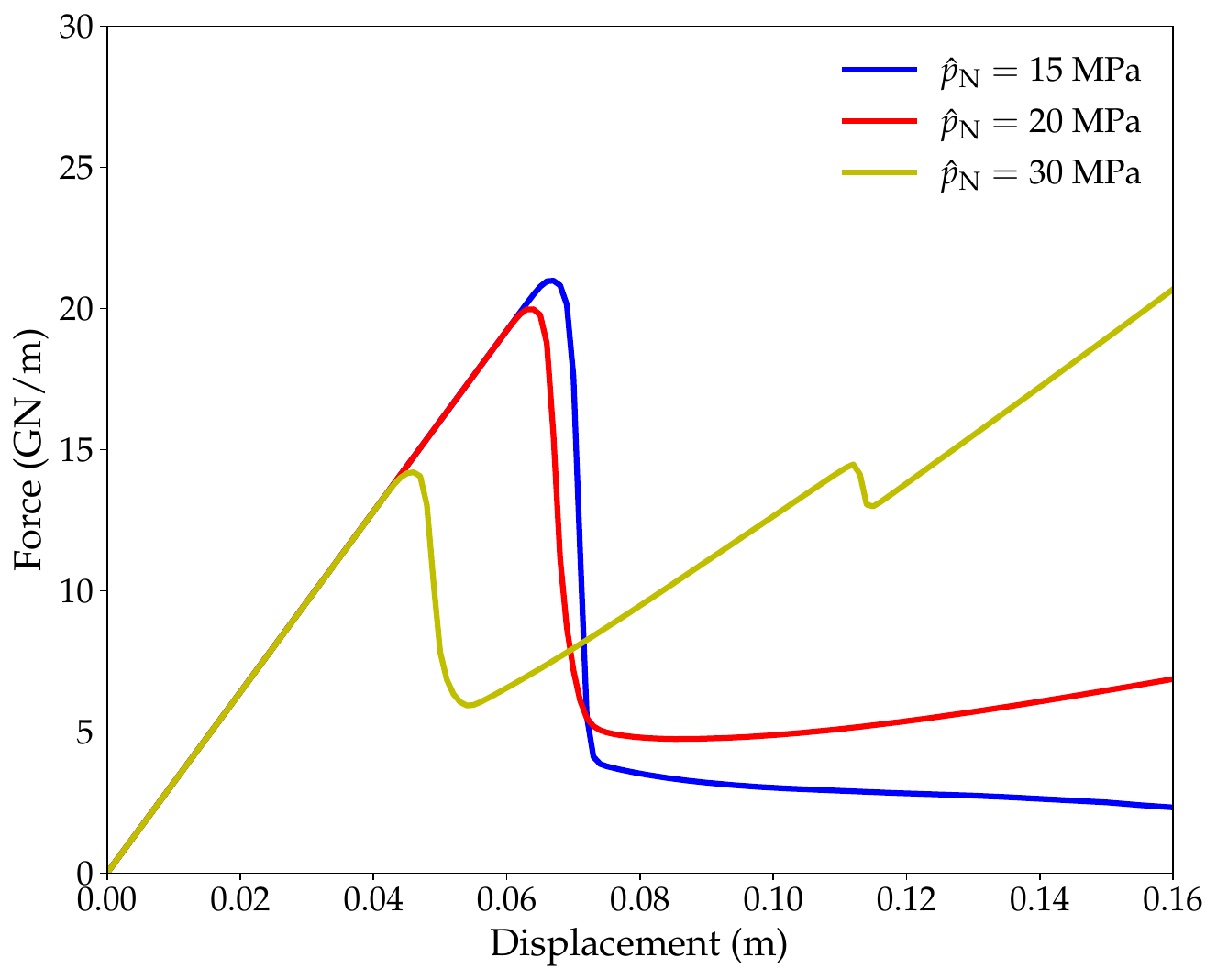} \label{fig:fault-growth-stressCompare-force}}
    \caption{Nucleation and propagation of faults from a weak zone: (a) ultimate phase-field distribution and (b) force-displacement curves for cases with different $\hat{p}_\mathrm{N}$.}
    \label{fig:fault-growth-stressCompare}
\end{figure}

\begin{figure}[h!]
    \centering
    \includegraphics[width=0.75\textwidth]{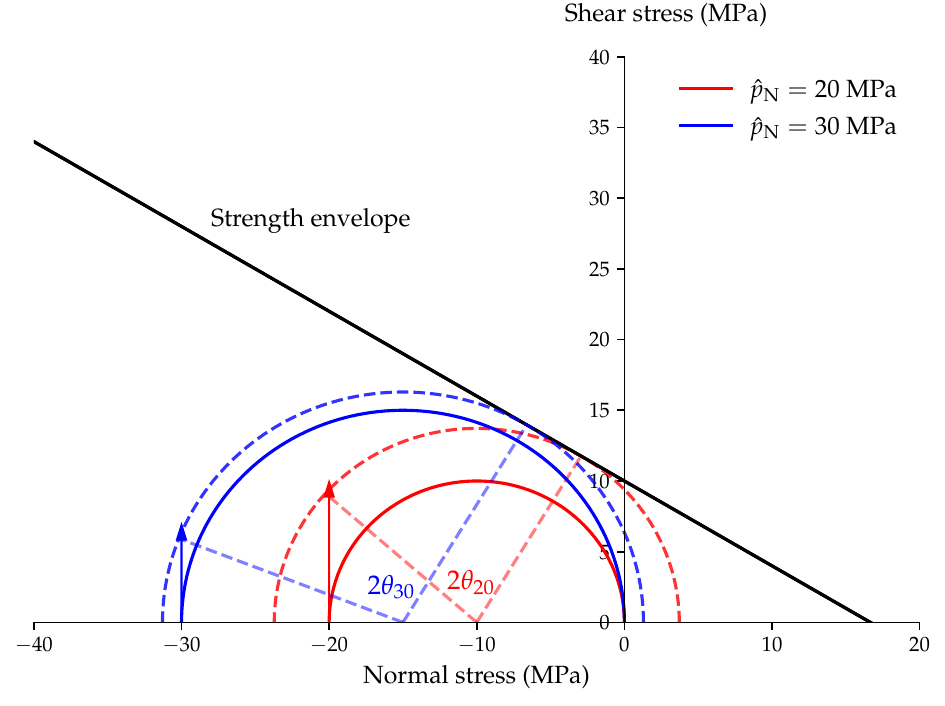}
    \caption{Nucleation and propagation of faults from a weak zone: Mohr's circles for $\hat{p}_\mathrm{N} = 20$ MPa and 30 MPa illustrating the effects of applied normal pressure $\hat{p}_\mathrm{N}$. Solid circles indicate the initial stress state for each case, while dashed circles represent the stress state when the peak shear strength is reached. Theoretical inclinations angles of slip surfaces are denoted by $\theta_{20}$ and $\theta_{30}$ for $\hat{p}_\mathrm{N} = 20$ MPa and 30 MPa, respectively. The length of the arrows indicates the required shear loading to reach the peak shear strength in each case.}
    \label{fig:mohr-circle}
\end{figure}

Finally, we investigate the effect of applied normal pressure $\hat{p}_\mathrm{N}$ on the phase-field results by repeating the simulation with $\hat{p}_\mathrm{N} = 15$ MPa and 30 MPa. 
As shown in Fig~\ref{fig:fault-growth-stressCompare-damage},  the final configuration of fault slip surfaces exhibits noticeable variations with changes in $\hat{p}_\mathrm{N}$. 
Specifically, the slip surfaces deviate further from the shear direction as the normal pressure increases.
For $\hat{p}_\mathrm{N} = 30$ MPa, multiple distributed slip surfaces form with increasing inclination angles, whereas for $\hat{p}_\mathrm{N} = 15$ MPa, a single, nearly horizontal slip surface spans the entire domain. 
This effect of $\hat{p}_\mathrm{N}$ on slip surface inclination can be explained using Mohr's circles, as illustrated in Fig.~\ref{fig:mohr-circle}.
Here, solid circles represent the initial stress states for each case, while dashed circles depict the stress state at the onset of slip surface growth (\ie~the peak shear strength is reached). 
For clarity, only the cases with $\hat{p}_\mathrm{N} = 20$ MPa and 30 MPa are shown. 
Using the Mohr's circle, the theoretical inclination angle between the dominant slip surface (\ie~R$_{1}$ shear) and the shear direction can be calculated, as denoted by $\theta_{20}$ and $\theta_{30}$ for $\hat{p}_\mathrm{N} = 20$ MPa and 30 MPa, respectively.   
Clearly, the case with $\hat{p}_\mathrm{N} = 30$ MPa has a higher inclination angle than the case with $\hat{p}_\mathrm{N} = 20$ MPa, consistent with the trends observed in the phase-field simulations.
It is worth noting that the actual inclination angles obtained from the simulations may deviate from the theoretical values due to non-ideal boundary conditions and the finite size of the weak zone.
However, the overall trend remains consistent: The inclination angles of the slip surfaces increase with higher $\hat{p}_\mathrm{N}$. 

In addition to influencing the slip surface configuration, normal pressure $\hat{p}_\mathrm{N}$ also affects the force-displacement curves, as shown in Fig.~\ref{fig:fault-growth-stressCompare-force}. 
Notably, the initial peak force decreases as $\hat{p}_\mathrm{N}$ increases, which may appear counterintuitive since normal confining pressure typically strengthens geologic materials. 
However, as illustrated by the Mohr's circles in Fig.~\ref{fig:mohr-circle}, less shear stress (represented by the length of the arrow) is required to reach the peak shear strength at higher $\hat{p}_\mathrm{N}$, thereby validating the phase-field simulation results.
Additionally, the case with $\hat{p}_\mathrm{N} = 15$ MPa exhibits a gradual decay of shear force to a residual value in the post-peak regime. 
In contrast, for $\hat{p}_\mathrm{N}=30$ MPa, the shear force demonstrates continuous hardening even beyond the peak force. 
This behavior can be explained as follows. 
For $\hat{p}_\mathrm{N}$ = 15 MPa, the single, flat slip surface (Fig.~\ref{fig:fault-growth-stressCompare-damage}) directly connects the lateral boundaries where no displacement constraints are applied. 
This penetrating slip surface acts as a weak plane, accommodating most of the deformation during shear and facilitating horizontal movement of the domain. 
Conversely, at $\hat{p}_\mathrm{N}$ = 30 MPa, the slip surfaces are more oblique to the shear direction and connect the top and bottom boundaries, which are fully constrained. 
As a result, despite the presence of multiple slip surfaces, most deformation is borne by the bulk material, leading to overall hardening behavior.
A similar transition in force-displacement behavior was observed in a compactivity-based plasticity model by Ma and Elbanna~\cite{ma2018strain}, where strain hardening at high $\hat{p}_\mathrm{N}$ was attributed to ductile behavior caused by less localized deformation in the gouge layer.
Here, the phase-field results exhibit consistent features, showing that deformation is more distributed within the bulk material and less localized along the slip surfaces at higher $\hat{p}_\mathrm{N}$.
Thus, the proposed phase-field model demonstrates potential as a promising tool for understanding faulting processes and the resulting mechanical behaviors in gouge materials. 

\subsection{Nucleation, propagation, and coalescence of faults in a stepover model}
\begin{figure}
    \centering
    \includegraphics[width=\textwidth]{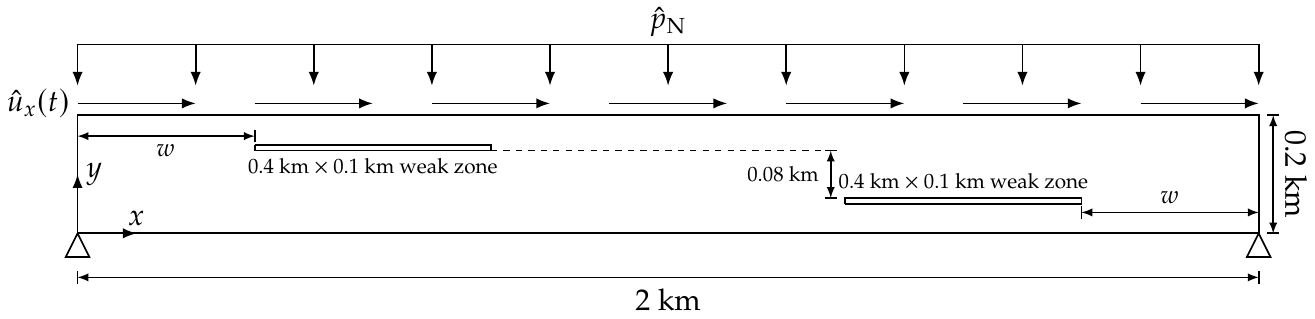}
    \caption{Nucleation, propagation, and coalescence of faults from two strips with stepover: problem setup.}
    \label{fig:fault-strips-setup}
\end{figure}

In the second example, we apply the phase-field model to a problem with more complex structural properties.
Specifically, we modify the setup from Section 4.2, by replacing the circular weak zone with two strips of weak zones separated by a 0.08 km stepover, as shown in Fig.~\ref{fig:fault-strips-setup}. 
These narrow strips are assigned a lower critical fracture energy, $\mathcal{G}_{II} = 0.1$ MJ/m$^2$, to promote the nucleation of slip surfaces.
The distance from each strip to the nearest lateral boundary of the domain is denoted by $w$. 
Two cases with different $w$ values are simulated: (i) $w = 0.3$ km, corresponding to two non-overlapping strips, and (ii) $w = 0.7$ km, corresponding to two overlapping strips. 
Material properties remain the same as in the previous example. 
The regularization length, element size, and time step are also unchanged, with $L=0.004$ km, $L/h = 5$, and $\Delta t = 0.05$ s, respectively.

\begin{figure}[htbp]
    \centering
    \subfloat[]{\includegraphics[width=0.53\textwidth]{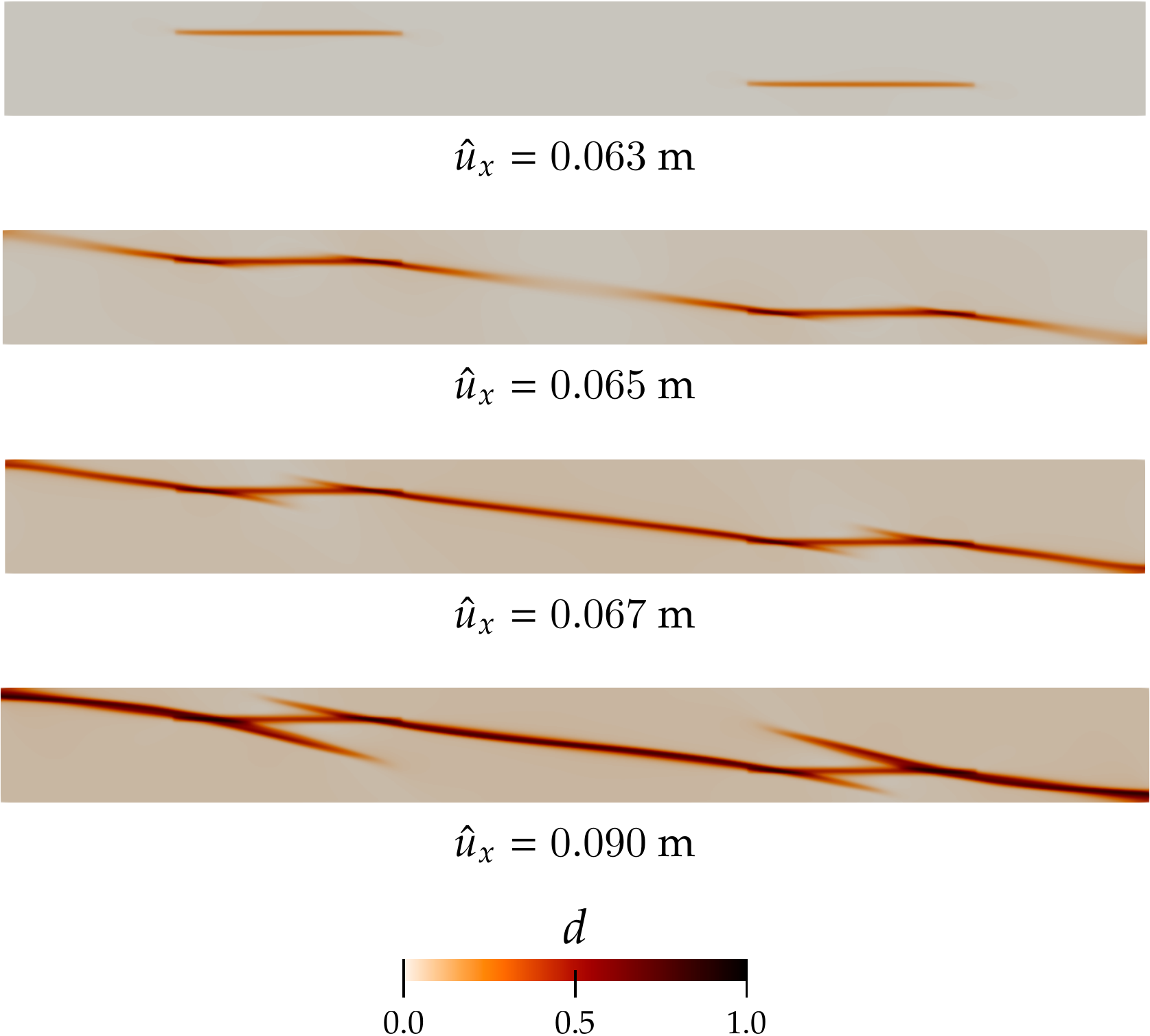}}
    \hspace{0.03\textwidth}
    \subfloat[]{\includegraphics[width=0.43\textwidth]{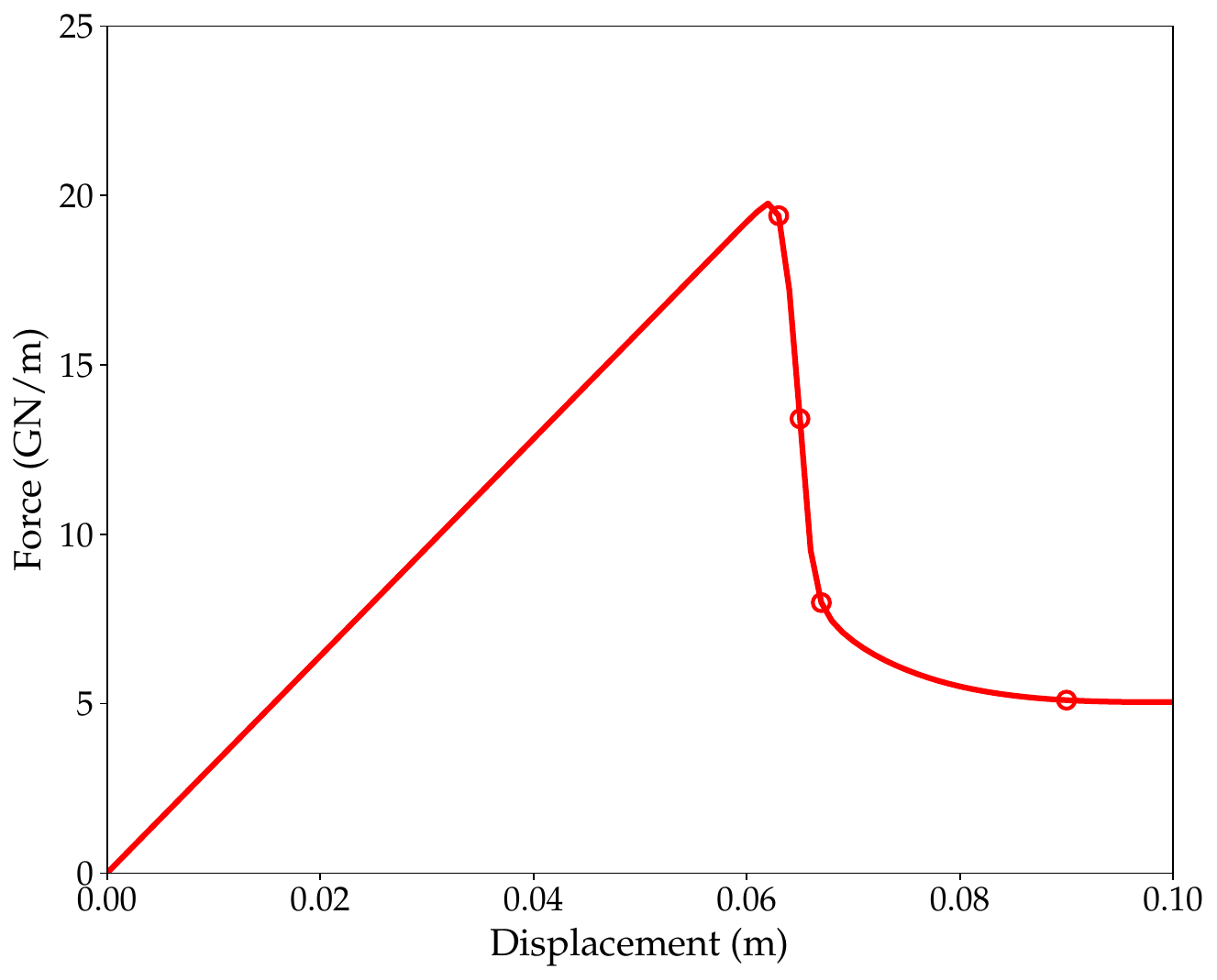}}
    \caption{Nucleation, propagation, and coalescence of faults from two strips with stepover: (a) phase-field evolution and (b) force-displacement curve for the case with $w = 0.3$ km (non-overlapping strips).}
    \label{fig:fault-strips-nonoverlap}
\end{figure}

Figure~\ref{fig:fault-strips-nonoverlap} illustrates the growth of slip surfaces for the case where the two strips have no overlap, with each stage of evolution highlighted on the force-displacement curve. 
At the peak force, initial slip surfaces form at the weak strips, creating a non-overstepping en \'{e}chelon pattern.
Subsequently, in the post-peak stage, these slip surfaces propagate outward from their tips. 
As shear continues, the two surfaces gradually coalesce at the center of the domain, forming a continuous slip surface that spans the entire layer. 
Concurrently, the shear force decreases and converges to a residual value. 
This modeled slip surface structure qualitatively aligns with observations from both experimental tests~\cite{cheng2015experimental} and other numerical simulations~\cite{wong2019recognition}, which show that substantial shearing drives the propagation of non-overlapping en \'{e}chelon fractures from their tips, eventually linking them together.

\begin{figure}[htbp]
    \centering
    \subfloat[]{\includegraphics[width=0.53\textwidth]{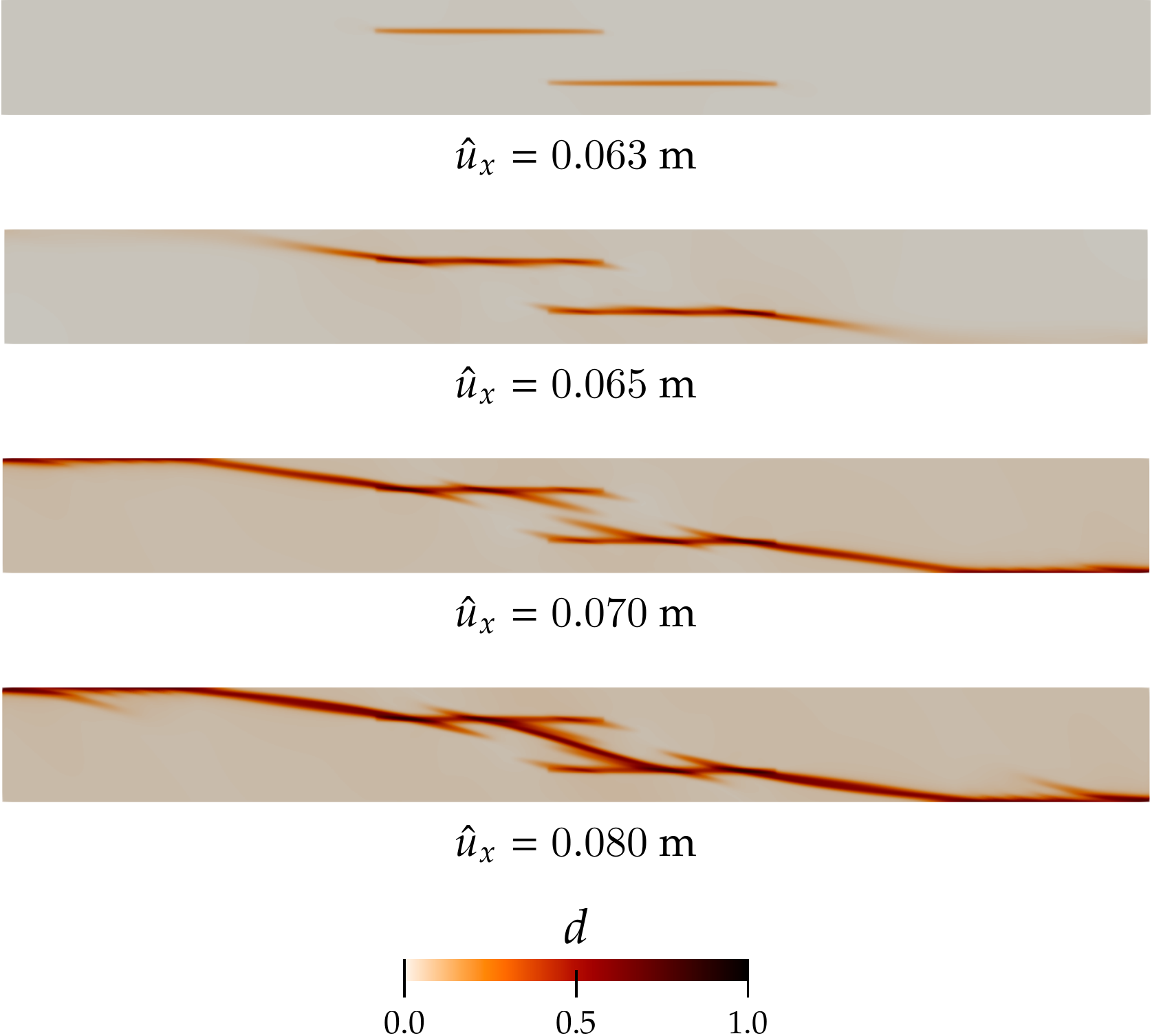} \label{fig:fault-strips-overlap-damage}}
    \hspace{0.03\textwidth}
    \subfloat[]{\includegraphics[width=0.43\textwidth]{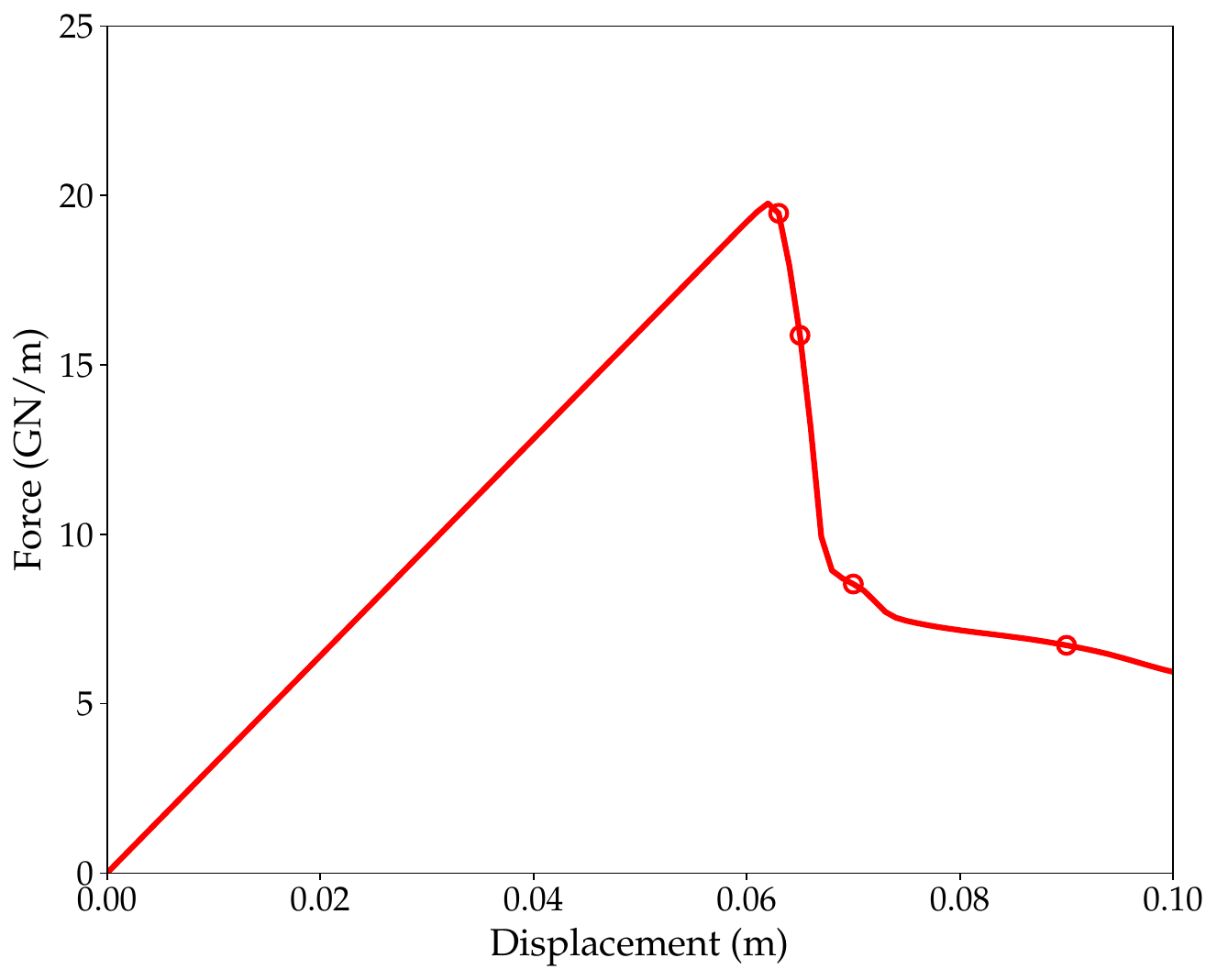} \label{fig:fault-strips-overlap-force-disp}}
    \caption{Nucleation, propagation, and coalescence of faults from two strips with stepover: (a) phase-field evolution and (b) force-displacement curve, for the case with $w = 0.7$ km (overlapping strips).}
    \label{fig:fault-strips-overlap}
\end{figure}

Figure~\ref{fig:fault-strips-overlap-damage} illustrates the nucleation and propagation of fault slip surfaces for the overlapping case, corresponding to the displacements marked on the force-displacement curve in Fig.~\ref{fig:fault-strips-overlap-force-disp}.
The initial nucleation process is similar to the non-overlapping case: two initial slip surfaces form at the locations of the strips when the shear force reaches its peak. However, unlike the non-overlapping case, fault coalescence occurs through new slip surfaces propagating from the centers of the initial slip surfaces, rather than from their tips.
As the fault coalescence progresses and forms a continuous slip surface across the domain, the shear force gradually decreases and approaches a residual value (approximately 5 MPa), similar to that observed in the previous case.

The simulations of both non-overlapping and overlapping cases reveal a single integrated fault slip surface spanning the entire domain, albeit with distinct patterns. 
In the non-overlapping case, a kink forms to connect the initially distinct fault surfaces. In contrast, the overlapping case exhibits one or more secondary faults emerging within or along the boundaries of the overlapping zone.
Despite these structural differences, the mechanical responses, as reflected in the force-displacement curves, show little variation between the two cases. 
These findings suggest that, under substantial shearing, the macroscopic mechanical behavior of the fault zone tends to coalesce and homogenize, becoming largely independent of internal structural complexities. 
However, these complexities could play a significant role in near-fault variations in stress and material properties and might influence the seismic wavefield differently in a fully dynamic model.
Notably, our results align with numerous field and experimental observations indicating that fault zones may evolve toward systems with progressive weakening and reduced segmentation~\cite{ben2003characterization}. 
While this numerical study is preliminary, we believe that further investigation using the developed phase-field method holds great potential for providing deeper insights into the critical problem of fault zone maturation.

\subsection{Nucleation and propagation of faults in a heterogeneous domain}
\begin{figure}
    \centering
    \includegraphics[width=\textwidth]{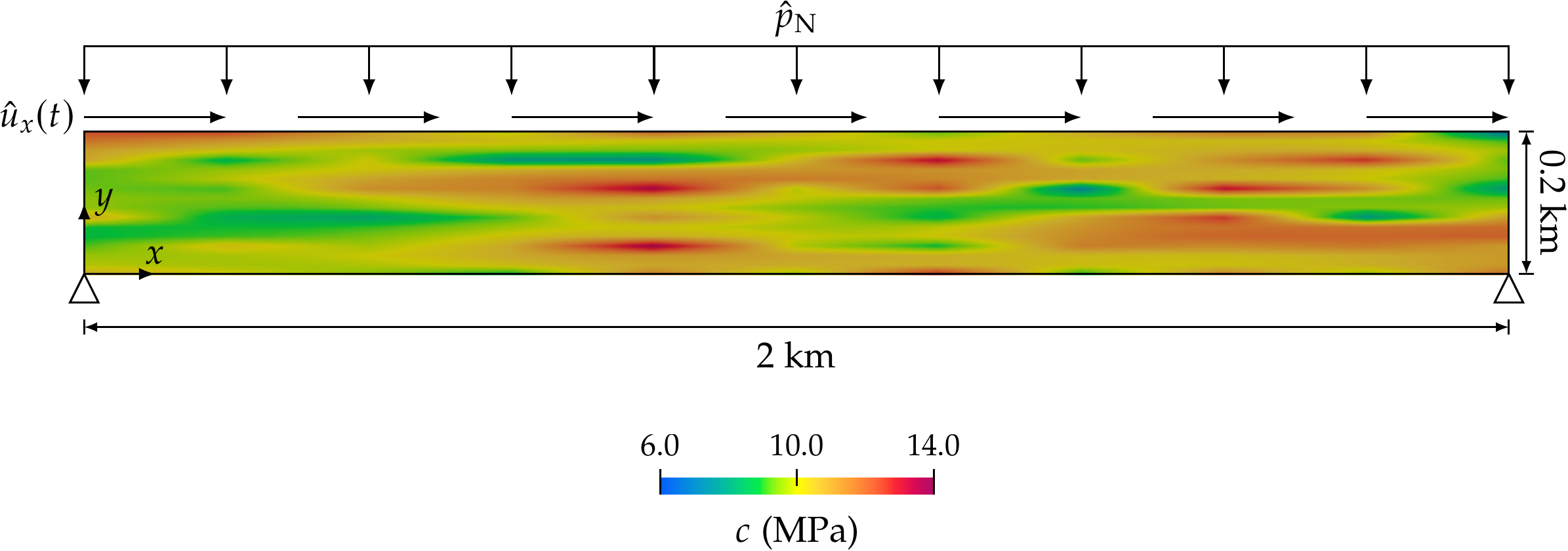}
    \caption{Nucleation and propagation of faults in a heterogeneous domain: problem setup.}
    \label{fig:fault-random-setup}
\end{figure}

In our final example, we simulate the nucleation and propagation of fault slip surfaces in a heterogeneous domain. A random field of cohesion is introduced, with values ranging from 6 MPa to 14 MPa.
Figure~\ref{fig:fault-random-setup} shows the specific problem setup and the spatial distribution of cohesion $c$ across the domain.
The domain size and boundary conditions are consistent with those used in the previous examples, but the weak zone is removed. The regularization length, spatial and temporal discretization levels, and material parameters (other than cohesion) remain unchanged.
The normal pressure applied to the top boundary is set to 20 MPa. 

\begin{figure}[htbp]
    \centering
    \subfloat[]{\includegraphics[width=0.53\textwidth]{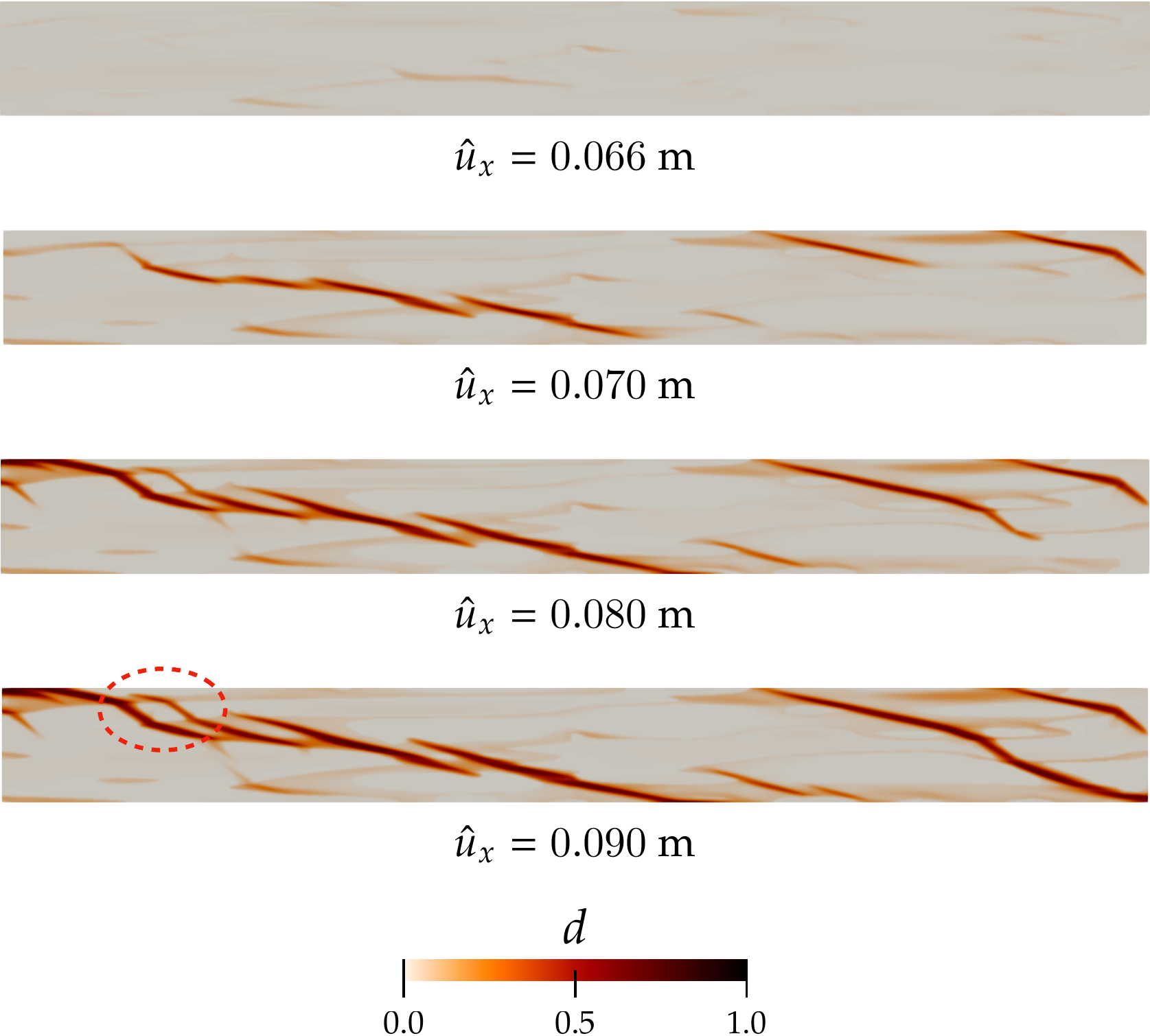}\label{fig:fault-random-20MPa-damage}}
    \hspace{0.03\textwidth}
    \subfloat[]{\includegraphics[width=0.43\textwidth]{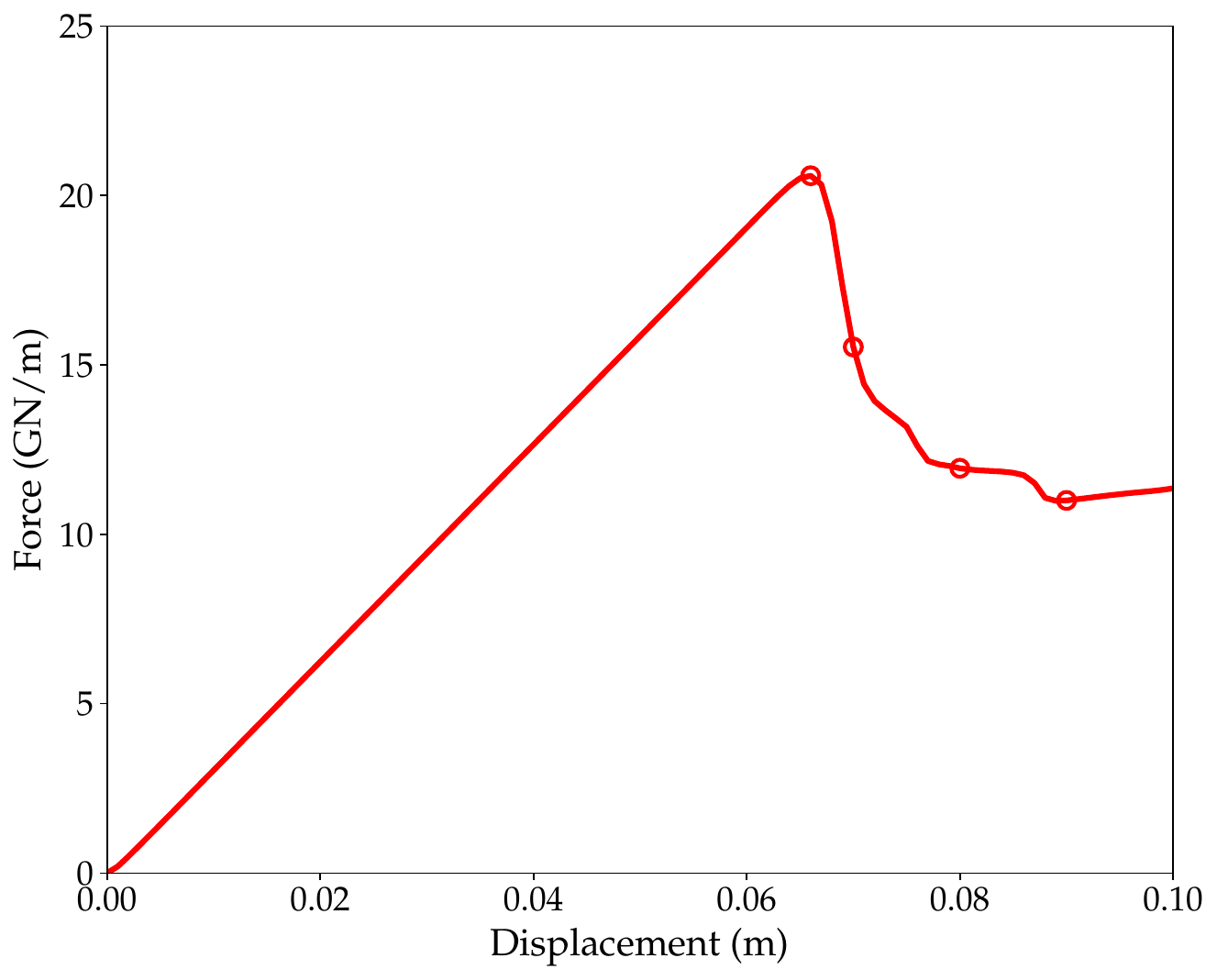}\label{fig:fault-random-20MPa-force-disp}}
    \caption{Nucleation and propagation of faults in a heterogeneous domain: (a) phase-field evolution and (b) force-displacement curve for the case with $\hat{p}_\mathrm{N} = 20$ MPa.}
    \label{fig:fault-random-20MPa}
\end{figure}

The evolution of fault slip surfaces in a heterogeneous domain is shown in Fig.~\ref{fig:fault-random-20MPa}, along with the force-displacement curve indicating the corresponding shear displacement at each growth stage. 
Initially, the shear force increases linearly with no observable damage growth.
At $\hat{u}_{x} = 0.066$ m, the shear force reaches its peak, coinciding with the emergence of diffuse damage in regions of low cohesion. 
In the post-peak stage, strain localization progressively accumulates in these initial damage zones, eventually forming fault slip surfaces. 
Compared to the results in Fig.~\ref{fig:fault-growth-damage-20MPa}, the slip surfaces generated in the heterogeneous domain display a more random distribution and intricate geometric features, such as kinking, as highlighted in Fig.~\ref{fig:fault-random-20MPa-damage}. 
Additionally, as evident in Fig.~\ref{fig:fault-random-20MPa-force-disp}, the softening behavior in the force-displacement curve becomes non-smooth.
These findings highlight the capability of the phase-field model to effectively handle complex fault geometries and mechanical responses induced by material heterogeneities.

% SECTION 6
% ------------------------------------------------------------------------------
\section{Closure}
\label{sec:closure}

In this paper, we have developed an advanced in-plane phase-field formulation for the quasi-dynamic modeling of fault rupture nucleation and propagation. 
Building on the previous anti-plane formulation, this in-plane version incorporates enhanced kinematics and the coupling between normal and shear stress components, enabling the modeling of rate- and state-dependent frictional slip and fault propagation.
The fault propagation direction criterion has been revised to account for variations in contact normal stress under in-plane conditions. 
Through numerical verification, we demonstrated that the phase-field results converge to the solutions of a discontinuous approach for in-plane modeling of quasi-dynamic, rate- and state-dependent frictional fault rupture.

Following verification, the phase-field model was applied to simulate fault nucleation and propagation under in-plane loading conditions. 
The phase-field results demonstrated consistent patterns of fault slip surfaces when compared with observations from relevant laboratory experiments and other numerical simulations, thereby validating the proposed formulation.
Additional simulations incorporating structural complexities and material heterogeneities further highlighted the exceptional capability of the phase-field method in capturing intricate fracture geometries arising from fault rupture and growth. 
Consequently, this model serves as a valuable tool for investigating the effects of complex structural and material features within fault cores, particularly on seismic wave propagation and velocity.
An important direction for future work is to extend the model to include fully dynamic computations, enabling explicit simulation of seismic wave propagation.

%----------------------------------------------------------------------------------------
% ACKNOWLEDGEMENTS
%----------------------------------------------------------------------------------------
% \acknowledgments
\section*{Acknowledgments}
This work was supported by the National Research Foundation of Korea (NRF) grant funded by the Korean government (MSIT) (No. RS-2023-00209799).
Portions of this work were performed under the auspices of the U.S. Department of Energy by Lawrence Livermore National Laboratory under Contract DE-AC52-07NA27344.

%----------------------------------------------------------------------------------------
% DATA AVAILABILITY STATEMENT
%----------------------------------------------------------------------------------------
\section*{Data availability statement}
The data that support the findings of this study are available from the corresponding author upon reasonable request.

% REFERENCES
% ------------------------------------------------------------------------------
\bibliography{references}

% APPENDIX
% ------------------------------------------------------------------------------
\appendix
\section{Discretization and algorithm for numerical solution}
\label{sec:appendix}

\subsection{Finite element discretization}
The standard Galerkin finite element method is employed to discretize the governing equations~\eqref{eq:momentum-balance-final} and~\eqref{eq:pf-evolution-final}. 
We define the trial solution spaces for the displacement field $\tensor{u}$ and the phase field $d$ as
\begin{align}
	\mathcal{S}_{u} &:= \left\{ \tensor{u} \mid \tensor{u} \in H^{1}, \:\: \tensor{u} = \hat{\tensor{u}} \:\: \text{on} \, \, \pd_{u} \Omega \right\}, \\
	\mathcal{S}_{d} &:= \left\{ d \mid d \in H^{1} \right\},
\end{align}
where $H^{1}$ denotes the Sobolev space of order one. 
The corresponding weighting function spaces are
\begin{align}
	\mathcal{V}_{u} &:= \left\{ \tensor{\eta} \mid \tensor{\eta} \in H^{1}, \:\: \tensor{\eta} = {\tensor{0}} \:\: \text{on} \, \, \pd_{u} \Omega \right\}, \\
	\mathcal{V}_{d} &:= \left\{ \phi \mid \phi \in H^{1} \right\}. 
\end{align}
Using the weighted residual method, the variational forms of the governing equations are expressed as
\begin{align}
	\mathcal{R}_{u} &:= - \int_{\Omega} \grad^{s} \tensor{\eta} : \tstress \: \dd V + \int_{\pd_{t} \Omega} \tensor{\eta} \cdot \hat{\tensor{t}} \: \dd A = 0, \label{eq:residual-disp} \\ 
	\mathcal{R}_{d} &:= \int_{\Omega} \phi g'(d) \mathcal{H}^{+} \: \dd V + \int_{\Omega} \dfrac{3 \mathcal{G}_{c}}{8L}\left(2L^2 \grad \phi \cdot  \grad  d + \phi \right) \: \dd V = 0 . \label{eq:residual-damage}
\end{align}
These residual equations are discretized spatially using finite elements; further details are omitted for brevity.

To solve the discretized equations, we make use of a staggered algorithm widely used in the phase-field literature~\cite{miehe2010phase}. Specifically, at each time step, we first solve Eq.~\eqref{eq:residual-disp} for the displacement field $\tensor{u}$ while holding $d$ fixed. Then, using the updated $\tensor{u}$, evaluate the crack driving force $\mathcal{H}^{+}$ and solve Eq.~\eqref{eq:residual-damage} for $d$.
This staggered approach is robust and achieves high accuracy with a single iteration per step, provided the time step size is sufficiently small.

Both nonlinear equations~\eqref{eq:residual-disp} and~\eqref{eq:residual-damage} are solved iteratively using the Newton--Raphson method. 
The linearized residual equations for the displacement field and the phase field are
\begin{align}
    \delta \mathcal{R}^{h}_{u} &:= \int_{\Omega} \grad^s \tensor{\eta}^{h} : \mathbb{C} : \grad^s \delta \tensor{u}^{h} \: \dd V, \label{eq:linearize-residual-disp}\\
    \delta \mathcal{R}^{h}_{d} &:= \int_{\Omega} \phi^{h} g''(d) \mathcal{H}^{+} \delta d^{h} \: \dd V + \int_{\Omega} \dfrac{3\mathcal{G}_{c}}{4}L \grad \phi^{h} \cdot \grad \delta d^{h} \: \dd V . \label{eq:linearize-residual-damage}
\end{align}
Here, $\delta$ denotes the linearization operator, $(\cdot)^{h}$ denotes the spatially discretized quantity, and $\mathbb{C}$ is the stress-strain tangent operator.
Notably, $\mathcal{H}^{+}$ is not linearized during the phase-field solution stage.

\subsection{Material update algorithm}

Algorithm~\ref{alg:material-update} details the procedure for updating internal variables, including the stress tensor, tangent operator, and crack driving force. 
Quantities at time $t_{n}$ are denoted with subscript $(\cdot)_{n}$, while those at $t_{n+1}$ have no subscript for brevity.
Several points in this algorithm may deserve elaboration. 
First, for intact material ($d=0$), the slip rate is undefined, and the friction coefficient is evaluated using the original Dieterich--Ruina law~\eqref{eq:rs-friction-classic}, giving $\mu = \mu_{0}$.
Second, for updating the state variable, we use the implicit Euler method and discretize the aging law~\eqref{eq:aging-law} as
\begin{align}
	\dot{\theta} = \dfrac{\theta - \theta_{n}}{\Delta t} = 1 - \dfrac{V \theta}{D_{c}} \:\: \text{with} \:\: \Delta t := t_{n+1} - t_{n} .
    \label{eq:discrete-aging-law}
\end{align}
Here, the slip rate is computed as 
\begin{align}
	V := \dot{\zeta_{f}} = \dfrac{\zeta_{f} - (\zeta_{f})_{n}}{\Delta t} = \dfrac{\Delta \zeta_{f}}{\Delta t}.  
	\label{eq:discrete-slip-rate}
\end{align}
Rearranging Eq.~\eqref{eq:discrete-aging-law}, we get 
\begin{align}
	\theta = \dfrac{\theta_{n} + \Delta t}{1+ \Delta t V/D_{c}}. \label{eq:return-mapping-theta}
\end{align}
Lastly, for the friction coefficient and degradation function, we assume identical peak and residual friction coefficients throughout the simulation for simplicity. 
For non-identical coefficients, we refer to Fei and Choo~\cite{fei2020phaseb}.

\begin{algorithm}
    \setstretch{1.25}
    \caption{Material point update algorithm for the phase-field formulation. }
    \label{alg:material-update}
    \begin{algorithmic}[1]
    \Require $\Delta t$, $\Delta \tstrain$, d, $\tensor{n}$, and $\tensor{m}$ at $t_{n+1}$. 
    \State Calculate $\tstrain = \tstrain_{n} + \Delta \tstrain$, $\Delta \gamma = 2 \Delta \tstrain : \tensor{\mathcal{S}}$, $\tstress_{m} = \mathbb{C}^{e}:\tstrain$, and $\tau_{m} = \tstress_{m}:\tensor{\mathcal{S}}$.
    \If{$d = 0$}
    \State Intact material. 
    \State Update $\bar{\tstrain } = \tstrain$, $\zeta_{f} = 0$, $\tstress = \tstress_{m}$, and $\mathbb{C} = \mathbb{C}^{e}$. 
    \State Update $\tau_{p} = p_{\mathrm{N}} \mu + c $ and $\tau_{r} = p_{\mathrm{N}} \mu $, where $\mu = \mu_{0}$ and $p_\mathrm{N} = -\tstress_{m}:(\tensor{n} \dyadic \tensor{n})$. 
    \If{$\lvert \tau_{m} \rvert > \tau_{p}$}
    \State Update $\mathcal{H}^{+} = \mathcal{H}_{t} + (\lvert \tau_{m} \rvert - \tau_{r})\lvert \Delta \gamma \rvert$. 
    \Else
    \State Keep $\mathcal{H}^{+} = \mathcal{H}_{t}$. 
    \EndIf
    \Else
    \State Compute the trial stress, $\tstress^\mathrm{tr}_{f} = \mathbb{C}^{e} : \bar{\tstrain}^\mathrm{tr}$ with $ \bar{\tstrain}^\mathrm{tr} = \left[\left(\bar{\tstrain} \right)_{n} +\Delta \tstrain \right]$.
    \State Update $\theta = (\theta_{n} + \Delta t )/(1 + \Delta t V_{n} /D_{c})$ and $\mu = \mu(V_{n}, \theta)$ according to Eq.~\eqref{eq:rs-friction-regularize}. 
    \State Calculate $\tau^\mathrm{tr}_{f} = \tstress^\mathrm{tr}_{f}: \tensor{\mathcal{S}}$ and evaluate $\alpha = \tau^\mathrm{tr}_{f}/\lvert \tau^\mathrm{tr}_{f} \rvert$.
    \State Perform return mapping to update $V$, $\theta$, $\zeta_{f}$, $\bar{\tstrain}$, $\mu$, $\tstress_{f}$, and $\mathbb{C}_{f}$. 
    \State Update $\tstress = g(d)\tstress_{m} + [ 1- g(d) ] \tstress_{f}$ and $\mathbb{C} = g(d) \mathbb{C}^{e} + [1 - g(d)] \mathbb{C}_{f}$. 
    \State Update $\mathcal{H}^{+} = \left(\mathcal{H}^{+}\right)_{n} + (\lvert \tau_{m} \rvert - \tau_{r} - \eta V)\lvert \Delta \gamma \rvert$, where $\tau_{r} = p_\mathrm{N} \mu$. 
    \EndIf
    \Ensure $\tstress$, $\mathbb{C}$, $\mathcal{H}^{+}$ at $t_{n+1}$. 
    \end{algorithmic}
\end{algorithm}

\subsection{Return mapping and tangent operator}

To complete the material update algorithm (Algorithm~\ref{alg:material-update}), we describe the return mapping procedure used to update the stress tensor and tangent operator in fractured material. The return mapping is performed in the full strain space, and it is detailed below.

The return mapping procedure begins with a vector of unknowns, comprising six independent strain components for continuous deformation and the increment of the slip magnitude
\begin{align}
    \tensor{x} = \left [ 
    \begin{array}{c}
        \left( \bar{\tstrain} \right)_{6 \times 1} \vspace{1em} \\
        \Delta \zeta_{f}
    \end{array}
    \right]_{7\times1} . 
\end{align}
The unknowns are calculated by solving the following residual equations
\begin{align}
    \tensor{r} = \left[ 
    \begin{array}{c}
        \left( \bar{\tstrain} - \bar{\tstrain}^\mathrm{tr} + \alpha \Delta \zeta_{f} \tensor{\mathcal{S}} \Gamma_{d}(d, \grad d) \right)_{6 \times 1} \vspace{1em} \\ 
        F
    \end{array}
    \right]_{7 \times 1} \rightarrow \tensor{0}. \label{eq:return-mapping-residual}
\end{align}
Here, the yield function is evaluated using $(p_\mathrm{N})_{n} := -(\tstress_{f})_{n}:(\tensor{n} \dyadic \tensor{n})$ as 
\begin{align}
    F(\tstress_{f}, V) = \lvert \tau_{f} \rvert - (p_\mathrm{N})_{n} \mu(V, \theta) - \eta V . \label{eq:return-mapping-yield-func}
\end{align}
This semi-implicit method greatly simplifies the formulation of the tangent operator and improves the robustness of the solution algorithm. 

The residual equations are solved using the Newton-Raphson method as
\begin{align}
    -\tensor{J} \cdot \Delta \tensor{x} = \tensor{r} \rightarrow \tensor{0} , 
\end{align}
where the Jacobian matrix is given by 
\begin{align}
    \tensor{J} = \left[ 
    \begin{array}{cc}
        \mathbb{I}_{6 \times 6} & \left(\alpha \tensor{\mathcal{S}} \Gamma_{d} (d, \grad d) \right)_{6 \times 1} \vspace{1em} \\  
        \left(\dfrac{\pd F}{\pd \tstress_{f}}: \mathbb{C}^{e} \right)^\intercal_{1 \times 6} & -\left((p_\mathrm{N})_{n}\dfrac{\dd \mu}{\dd V} + \eta \right) \dfrac{\pd V}{\pd \Delta \zeta_{f}}
    \end{array}
    \right]. 
\end{align}
Here, the derivative of the yield function~\eqref{eq:return-mapping-yield-func} with respect to $\tstress_{f}$ is given by 
\begin{align}
	\dfrac{\pd F}{\pd \tstress_{f}} = \dfrac{\tau_{f}}{\lvert \tau_{f} \rvert}\tensor{\mathcal{S}}. \label{eq:dF-dsigma}
\end{align}
The derivative of the slip rate with respect to the increment of the slip magnitude can be derived by taking the derivatives on both sides of Eq.~\eqref{eq:discrete-slip-rate}, which gives 
\begin{align}
  \dfrac{\pd V}{\pd \Delta \zeta_{f}} = \dfrac{1}{\Delta t} . 
\end{align}
For the derivative of the frictional coefficient with respect to the slip rate, we first apply the chain rule to $\dd \mu/\dd V$ as
\begin{align}
  \dfrac{\dd \mu}{\dd V} = \dfrac{\pd \mu}{\pd V} + \dfrac{\pd \mu}{\pd \theta} \dfrac{\pd \theta}{\pd V}. \label{eq:d-mu-d-v}
\end{align}
The regularized rate- and state-dependent friction formulation~\eqref{eq:rs-friction-regularize} gives 
\begin{align}
  \dfrac{\pd \mu}{\pd V} = \dfrac{aK}{\sqrt{1 + (KV)^{2}}} , \label{eq:pd-mu-pd-v}
\end{align}
and
\begin{align}
  \dfrac{\pd \mu}{\pd \theta} = \dfrac{bKV}{\theta\sqrt{1 + (KV)^{2}}}, \label{eq:pd-mu-pd-theta}
\end{align}
where
\begin{align}
  K = \dfrac{1}{2V_{0}} \exp \left [\dfrac{\mu_{0} + b \ln(V_{0}\theta/D_{c})}{a} \right] . 
\end{align}
The last term $\pd \theta/\pd V$ in Eq.~\eqref{eq:d-mu-d-v} can be obtained by taking the derivative with respect to the slip rate on both sides of Eq.~\eqref{eq:return-mapping-theta} 
\begin{align}
  \dfrac{\pd \theta}{\pd V} = - \dfrac{D_{c}\Delta t (\theta_{n} + \Delta t)}{(D_{c} + \Delta t V)^2} .  \label{eq:pd-theta-pd-v}
\end{align}
Inserting Eqs.~\eqref{eq:pd-mu-pd-v}, \eqref{eq:pd-mu-pd-theta}, and \eqref{eq:pd-theta-pd-v} into Eq.~\eqref{eq:d-mu-d-v}, we obtain the full expression of $\dd \mu/\dd V$ as
\begin{align}
  \dfrac{\dd \mu}{\dd V} = \dfrac{aK}{\sqrt{1 + (KV)^{2}}} - \dfrac{bKV}{\theta\sqrt{1 + (KV)^{2}}}\dfrac{D_{c}\Delta t (\theta_{n} + \Delta t)}{(D_{c} + \Delta t V)^2} . \label{eq:d-mu-d-v-final}
\end{align}

Finally, we evaluate the tangent operator in the fractured region, defined as 
\begin{align}
	\mathbb{C}_{f} = \dfrac{\pd \tstress_{f}}{ \pd \tstrain}.
\end{align}
Linearizing the residual equation~\eqref{eq:return-mapping-residual} yields
\begin{align}
    \left [ 
    \begin{array}{c}
        \left(\delta \bar{\tstrain} + \alpha \tensor{\mathcal{S}} \Gamma_{d}(d,\grad d) \delta \Delta \zeta_{f} \right)_{6 \times 1} \vspace{1em} \\
         \dfrac{\pd F}{\pd \tstress_{f}}: \mathbb{C}^{e}:\delta \bar{\tstrain} - \left((p_\mathrm{N})_{n}\dfrac{\dd \mu}{\dd V} + \eta \right) \dfrac{1}{\Delta t} \delta \Delta \zeta_{f}
    \end{array}
    \right] = 
    \left [ 
    \begin{array}{c}
        \left( \delta \bar{\tstrain}^\mathrm{tr} \right)_{6\times 1} \\
        0
    \end{array}
    \right]_{7 \times 1} .  \label{eq:linearize-return-mapping-residual}
\end{align}
Given $\bar{\tstrain}^\mathrm{tr} = \left[\left(\bar{\tstrain} \right)_{n} +\Delta \tstrain \right] = \left[\left(\bar{\tstrain} \right)_{n} + \tstrain - \tstrain_{n} \right] $ in Algorithm~\ref{alg:material-update}, we obtain
\begin{align}
	\delta \bar{\tstrain}^\mathrm{tr} = \delta \tstrain .
\end{align}
Inserting the above equation and $\delta\bar{\tstrain} := (\mathbb{C}^{e})^{-1}:\delta \tstress_{f}$ into Eq.~\eqref{eq:linearize-return-mapping-residual}, and replacing $\pd F/\pd \tstress_{f}$ with Eq.~\eqref{eq:dF-dsigma} lead to the following two separate equations 
\begin{align}
	\left(\mathbb{C}^{e}\right)^{-1} : \delta \tstress_{f} + \alpha \tensor{\mathcal{S}} \Gamma_{d}(d,\grad d) \delta \Delta \zeta_{f} = \delta \tstrain, \label{eq:linearize-strain-balance}
\end{align}
and 
\begin{align}
	\left( \dfrac{\tau_{f}}{\lvert \tau_{f} \rvert} \tensor{\mathcal{S}}\right): \delta \tstress_{f} - \left((p_\mathrm{N})_{n}\dfrac{\dd \mu}{\dd V} + \eta \right) \dfrac{1}{\Delta t} \delta \Delta \zeta_{f} = 0 . \label{eq:linearize-yield-func}
\end{align}
Rearranging Eq.~\eqref{eq:linearize-strain-balance} gives
\begin{align}
	\delta \tstress_{f} = \mathbb{C}^{e}:\left[\delta \tstrain - \alpha \tensor{\mathcal{S}} \Gamma_{d}(d,\grad d) \delta \Delta \zeta_{f} \right] . \label{eq:delta-stressf}
\end{align}
Differentiating Eq.~\eqref{eq:delta-stressf} with respect to the total strain $\tstrain$ gives the tangent operator in fractured material as
\begin{align}
	\mathbb{C}_{f} := \dfrac{\pd \tstress_{f}}{\pd \tstrain} =  \mathbb{C}^{e}:\left[\mathbb{I} - \alpha  \Gamma_{d}(d,\grad d) \left(\tensor{\mathcal{S}}\dyadic  \dfrac{\pd \Delta \zeta_{f} }{\pd \tstrain} \right)\right] . \label{eq:cto-fracture}
\end{align}
To obtain the final expression of $\mathbb{C}_{f}$, we evaluate the last term in the above equation, $\pd \Delta \zeta_{f} / \pd \tstrain$.
For this, we substitute Eq.~\eqref{eq:delta-stressf} into Eq.~\eqref{eq:linearize-yield-func} and get
\begin{align}
	 \left( \dfrac{\tau_{f}}{\lvert \tau_{f} \rvert} \tensor{\mathcal{S}}\right): \mathbb{C}^{e}: \delta \tstrain = \left[ \dfrac{\tau_{f}}{\lvert \tau_{f} \rvert} \left( \tensor{\mathcal{S}}: \mathbb{C}^{e}: \tensor{\mathcal{S}} \right) \alpha \Gamma_{d}(d, \grad d) + \left((p_\mathrm{N})_{n}\dfrac{\dd \mu}{\dd V} + \eta \right) \dfrac{1}{\Delta t} \right]\delta \Delta \zeta_{f} . 
\end{align}
We then rearrange the above equation and differentiate it with respect to $\tstrain$, which gives
\begin{align}
	\dfrac{\pd \Delta \zeta_{f}}{ \pd \tstrain} = \dfrac{\dfrac{\tau_{f}}{\lvert \tau_{f} \rvert} \mathbb{C}^{e} : \tensor{\mathcal{S}}}{ \left[ \dfrac{\tau_{f}}{\lvert \tau_{f} \rvert} \left( \tensor{\mathcal{S}}: \mathbb{C}^{e}: \tensor{\mathcal{S}} \right) \alpha \Gamma_{d}(d, \grad d) + \left((p_\mathrm{N})_{n}\dfrac{\dd \mu}{\dd V} + \eta \right) \dfrac{1}{\Delta t} \right]} . 
\end{align}
Inserting the above equation into Eq.~\eqref{eq:cto-fracture} results in the final form of the tangent operator in fractured material as
\begin{align}
	\mathbb{C}_{f} = \mathbb{C}^{e} - \alpha \Gamma_{d}(d, \grad d) \dfrac{\tau_{f}}{\lvert \tau_{f} \rvert} \dfrac{ \left(\mathbb{C}^{e} : \tensor{\mathcal{S}} \right) \dyadic \left(\mathbb{C}^{e} : \tensor{\mathcal{S}} \right)}{ \left[ \dfrac{\tau_{f}}{\lvert \tau_{f} \rvert} \left( \tensor{\mathcal{S}}: \mathbb{C}^{e}: \tensor{\mathcal{S}} \right) \alpha \Gamma_{d}(d, \grad d) + \left((p_\mathrm{N})_{n}\dfrac{\dd \mu}{\dd V} + \eta \right) \dfrac{1}{\Delta t} \right]} .
\end{align}

\end{document}